    \documentclass[10pt]{article}

  \usepackage{graphicx}                 
  
\begin{document}

\title{Quantum Measurements: a modern view for quantum optics experimentalists}

\author{Aephraim M. Steinberg}

\maketitle

\begin{center}
\small{Centre for Quantum Information \& Quantum Control and Department of Physics,} \\
\small{University of Toronto, Toronto, Ontario, Canada; and}\\
\small{Canadian Institute for Advanced Research}
\end{center}


 \newcommand{\um}[1]{\"{#1}}
 \renewcommand{\Im}{{\protect\rm Im}}
 \renewcommand{\Re}{{\protect\rm Re}}
 \newcommand{\ket}[1]{\mbox{$|#1\protect\rangle$}}
 \newcommand{\bra}[1]{\mbox{$\protect\langle#1|$}}
 \newcommand{\proj}[1]{\mbox{$\ket{#1}\bra{#1}$}}
 \newcommand{\expect}[1]{\mbox{$\protect\langle #1 \protect\rangle$}}
 \newcommand{\inner}[2]{\mbox{$\protect\langle #1 | #2 
\protect\rangle$}}

%
%

\tableofcontents

\vspace{0.3in}

\small{Lectures given at the 101st Les Houches summer school, on ``Quantum Optics and Nanophotonics", August 2013}

\small{to be published by Oxford University Press}

%
%
%

\pagebreak
%

In these notes, based on lectures given as part of the Les Houches summer school on Quantum Optics and Nanophotonics in August, 2013, I have tried to give a brief survey of some important approaches and modern tendencies in quantum measurement.  I wish it to be clear from the outset that I shy explicitly away from the ``quantum measurement problem,'' and that the present treatment aims to elucidate the theory and practice of various ways in which measurements can, in light of quantum mechanics, be carried out; and various formalisms for describing them.  While the treatment is by necessity largely theoretical, the emphasis is meant to be on an experimental ``perspective'' on measurement -- that is, to place the priority on the possibility of gaining information through some process, and then attempting to model that process mathematically and consider its ramifications, rather than stressing a particular mathematical definition as the {\it sine qua non} of measurement.  The textbook definition of measurement as being a particular set of mathematical operations carried out on particular sorts of operators has been so well drilled into us that many have the unfortunate tendency of saying ``that experiment can't be described by projections onto the eigenstates of a Hermitian operator, so it is not really a measurement,'' when of course any practitioner of an experimental science such as physics should instead say ``that experiment allowed us to measure something, and if the standard theory of measurement does not describe it, the standard theory of measurement is incomplete.''  Idealisations are important, but when the real world breaks the approximations made in the theory, it is the theory which must be fixed, and not the real world.

\section{Information From Measurement: Probabilities and Update Rules}

The defining characteristic of measurement -- whether in classical or in quantum physics -- is that it increases our information about a system (or a process).   From the perspective which views the (e.g. quantum) state as no more than an expression of our information, it is therefore tautological that measurements disturb systems' states.  While the disturbances occasioned by quantum measurement are richer and perhaps more surprising than the trivial classical observation that a 50\% chance of rain ``collapses'' into either 100 or 0 once the fact of the matter is observed, much of their behaviour can be derived by first thinking about the how to modify one's description of a state upon successful gain of information via a measurement.  This is a well-trod topic in classical statistics, but I treat it (briefly) here for several reasons.  First, while such ``update rules'' in some sense are the fundamental basis of all experimental sciences, they are widely misunderstood by physicists, presumably because the approximate rules we learn in our laboratory courses suffice in so many cases.  Second, a clear perspective on the axioms of quantum measurement can be derived from these considerations, and leads naturally to some of the modern generalisations of the quantum measurement formalism which I will discuss here.  Finally, a great deal of work in recent years has focused on quantum state and process estimation (``tomography'') and on quantum metrology, and some controversies in those areas in fact stem from older questions in classical statistics.

\subsection{Classical information Update and Bayes's Rule}
\label{IA0}

Let us consider a classic example of parameter estimation, whose relationship to quantum measurement should become obvious.  Suppose we are given a coin with probabilities $p_H$ of coming up heads and $p_T=1-p_H$ of coming up tails.  Lacking prior knowledge of $p_H$, how would we estimate it?  If we flip the coin $N$ times, say that we find $H$ heads and $T=N-H$ tails.  Our intuition that $p_H$ is best estimated to be $H/N$ is well founded.  This is, in the technical sense of the term (which I shall discuss presently) the ``most likely'' solution.  But what uncertainty should we report on this estimate?  We know that a sequence of fair coin tosses with probability of heads equal to $p_H$ will lead to binomial statistics, that is, $p_H N \pm \sqrt{N (p_H)(1-p_H)}$, for an {\it observed} ratio of $p_H \pm \sqrt{p_H(1-p_H)/N}$.  Usual (which is to say careless, but oftentimes acceptable) practice is to report this latter quantity as the uncertainty in our result.  That is, we would report our finding as an estimate of $p_H$ which I will term $p_{\rm est} = H/N \pm \sqrt{(H/N)(1-H/N)/N}$.  The error in this approach can be seen easily by considering the case in which $0$ heads appeared.  Our ``most likely'' estimate of $p_H$ is indeed $0$, but the formula given above would yield an uncertainty of $0$.  Since it is {\it possible} for $N$ coin tosses to all come up tails, so long as $p_H \neq 1$, we would not wish to report $p_{\rm est} = 0 \pm 0$.  The mistake was to conflate the probability of getting $H$ heads {\it if the true probability were $p_{\rm est}$} and the probability {\it that $p_{\rm est}$ is equal to\footnote{or more carefully, within a specified confidence interval of} the true $p_H$} given that we obtained $H$ heads.  Mathematically, using the notation $P(A|B)$ for the probability of $A$, {\it given} that we know $B$:
\begin{eqnarray}
P(\;H=0\;|\;p_H=0\;) & = & 100\% \;  , \nonumber \\
{\rm yet} \;\;\;P(\;p_H=0\;|\;H=0\;) & \neq & 100 \% \; 
\end{eqnarray}
(because, for instance, $P(\;H=0\;|\;p_H=0.0001\;) \neq0$).  In experimental science, we use observed data to test and constrain models.  That is, based on our observations, we update our estimates of the probability that a given model is correct or incorrect; in many contexts (such as the present one), a ``model'' may simply be the value of a parameter.  (When Millikan reported a charge for the electron and an uncertainty, he was in essence reporting a probability distribution for a {\it parameter} in his theory.)  The trick is that our models allow us to make (generally probabilistic) predictions about what we would expect to observe, {\it given} particular values for all relevant parameters, $P({\rm data} \; | \; {\rm model})$; while what we wish to know if the probability of the model being correct (or the parameters having a given set of values), {\it given} our observed data, $P({\rm model} \; | \; {\rm data})$.  Fortunately, classical probability theory has a formula -- Bayes's Theorem -- for carrying out just this inversion.  Because a joint probability of two propositions $A$ and $B$ can be factorized in either of two ways -- $P(A\&B) = P(B) P(A|B) = P(A) P(B|A)$ --,
\begin{equation}
P(B|A) = \frac{P(A|B)P(B)}{P(A)} \; .
\label{Bayes}
\end{equation}

Applied to the classical coin example, we can write the probability that $p_H$ has some particular value, given that we observed $H$ to be $0$, as $P(p_H\;|\;H=0) = P(H=0\;|\;p_H)P(p_H)/P(H=0)$.  Note that $p_H$ here is a variable; the denominator on the right-hand side is independent of $p_H$ (which is a good thing, since after having observed that $H=0$, and without any other information about the model, how should we estimate what the prior probability was that we {\it would} have observed $H=0$?).  In other words, it merely serves as a normalisation constant.  Now, we already know how to calculate the probability of $H$ heads from the binomial distribution
\begin{equation}
P(H\;|\;p_H) = p_H^H(1-p_H)^{N-H} \left(
\begin{array}{c}
N \\
H \\
\end{array} \right) \; ,
\end{equation}
and since $\left(
\begin{array}{c}
N \\
H \\
\end{array} \right)$ is independent of $p_H$ and we will normalize our distribution in any event, all that concerns us is that $P(H\;|\;p_H) \propto p_H^H(1-p_H)^{N-H}$.

This function is termed the ``likelihood'' ${\cal L}(p_H) \equiv P(H\;|\;p_H)$ of the model.  Although in everyday speech, ``likelihood'' and ``probability'' may be used synonymously, it is essential to note that ${\cal L}$ is {\it not} the probability that your model $p_H$ is correct in light of your observations, but rather the probability that you {\it would} have obtained those observations {\it if} $p_H$ had been correct.  

A common estimation technique -- the one we implicitly used when we concluded that our best estimate of $p_H$ was $H/N$ -- is known as ``maximum likelihood estimation,'' for reasons which should be obvious.  It is a simple exercise to show that this likelihood function ${\cal L}(p_H) \propto p_H^H(1-p_H)^{N-H}$ attains its maximum when $p_H = H/N$.  But consider the case for $H=0$, sketched in figure 1.  Here the likelihood of the model is just the probability of flipping $N$ tails in a row: ${\cal L} = p_T^N$ where $p_T=1-p_H$ is the probability of flipping tails.  The point of maximum likelihood is $p_H = H/N = 0$, but there are two issues to address.  The first is whether or not we can even conclude that $p_H=0$ is the {\it most probable} value of $p_H$ in light of our observations.  Recall that from Eq.~\ref{Bayes}, $P(p_H|H)\propto {\cal L}(p_H)P(p_H)$ and not just ${\cal L}(p_H)$ on its own.  This $P(p_H)$ on the right-hand side is the {\it prior probability} (or probability density, in the case of a continuous parameter like $p_H$) that $p_H$ was true to begin with.  Being careless with these ``priors'' is one of the most common errors made with statistics.

\begin{figure}[tb]
\centering
\includegraphics[width=4in]{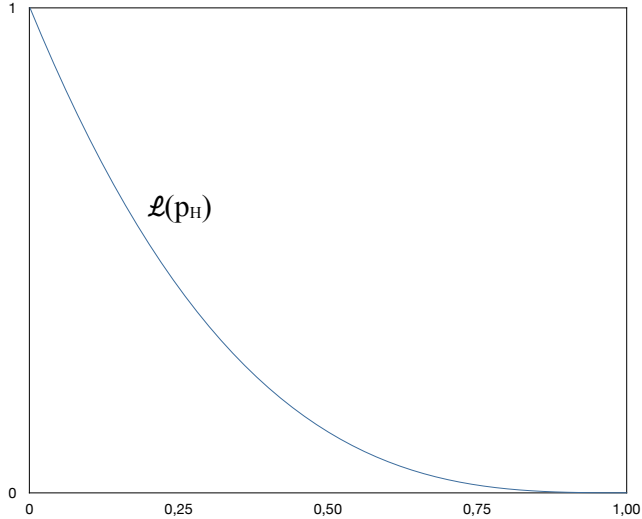}
\caption{The likelihood of a model $p_H$ for the case where $H=0$ heads are tossed (example shown for $N=3$)}
\label{fig1}
\end{figure}

A simple famous example is the following.  Suppose that you are tested for a rare disease (incidence $10^{-6}$).  The test is excellent; the probability of a positive result is $99\%$ if you have the disease, and only $0.01\%$ if you do not.  Suppose you take the test and it comes back positive for the disease -- which ``theory'' (``d,'' that you have the disease; or ``h,'' that you are healthy) has the maximum likelihood?  
\begin{eqnarray}
{\cal L}(d) \equiv P(+\;|\;d) & = & 99\% \nonumber \\
{\cal L}(h) \equiv P(+\;|\;h) & = & 0.01\% \; .
\end{eqnarray}
Clearly, the likelihood of ``d'' is nearly $10^4$ times larger than that of ``h.''  But which is more {\it probable}?  Well, since the test detects almost all true cases of the disease, roughly $10^{-6}$ of the population gets a `+' because they have the disease.  On the other hand, since almost all the population is healthy, and the probability of a false positive is $10^-4$, roughly $10^{-4}$ of the population gets a `+' because of simple error.  In other words, 100 times more people get `+'s by error than because of the presence of the disease; there is still a $99\%$ chance that you are healthy.  Of course, the $1\%$ chance that you have the disease is $10^4$ times larger than the $10^{-6}$ estimate you would have made before taking the test; in fact, the {\it increase} in this probability (more strictly, in the ratio of the two probabilities) is given precisely by the ratio of the likelihoods:
\begin{equation}
\left[ \frac{P(d|+)}{P(h|+)} \right]= \frac{{\cal L}(d)}{{\cal L}(h)} \; \left[\frac{P(d)}{P(h)}\right] \; .
\label{ratio}
\end{equation}
What this shows us is that our experimental observations on their own (from which we calculate the likelihoods of the various models) cannot tell us the probability of one model or another being correct.  The likelihoods serve only to {\it update} our estimates of these probabilities.  Sadly, in survey after survey, a majority of doctors do not solve these problems correctly.  Less importantly, but perhaps more shockingly, the same is true of a majority of physicists.

This interpretation of the likelihood based on data as an ``update rule'' is crucial when one wishes to combine information from multiple experiments or observations.  It is also important when other constraints exist for theoretical reasons -- for instance, early measurements of the square $m_\nu^2$ of the neutrino mass consistently turned up negative numbers, which were of course inconsistent with our understanding of physics.  Had these results occurred simply because of experimental uncertainties\footnote{in fact, systematic errors were eventually identified}, the correct interpretation would have been to say that even though the point of maximum likelihood occurred for $m_\nu^2<0$, the prior probability $P(m_\nu^2)$ vanished for negative squared-masses, and no amount of data would change this.  A reasonable candidate for $P(m_\nu^2 | {\rm data})$ would be truncated below $m_\nu^2=0$, leaving an asymmetric curve not unlike Figure 1.

But what should we do when we have no prior information, as in the case of the coin?  The natural assumption might be to assume a ``flat'' prior.  This in itself can be tricky; for instance, making $P(m^2)$ flat is different from making $P(m)$ flat.  Making $P({\rm log} m)$ flat is even more different.  This is the problem of choosing a measure.  Many therefore advocate using a scale-invariant prior, the ``Bures prior.''  For most applications in physics, this question is secondary.  The fact is that we never prove or disprove a theory based on a single observation, but rather on a large set of observations.  Every new observation multiplies the probability ratio in Eq.~\ref{ratio} by another data-based ratio of likelihoods; in the end, one hopes to be relatively insensitive to the initial choice of prior.

So, let us for simplicity suppose a flat prior, $P(p_H) dp_H = dp_H$ for $p_H$ between $0$ and $1$.  After flipping $N$ tails, we {\it can} now conclude that $P(p_H|{\rm data}) \propto {\cal L}(p_H) \propto p_T^N$.  But would we really wish to report as our best estimate the {\it peak} of this asymmetric function, at $p_H=0$, given that we know with certainty that the true value can not possibly lie below that?  Or wouldn't we rather report the mean of the distribution?  You are all familiar with least-squares fits; typically, our goal is to minimize the squared-error.  Which value of $p_H$ would we report in order to minimize the expectation value of the squared-error from the true value of $p_H$?  The answer is precisely the mean, $\expect{p_H} = \int_0^\infty dp_H \, p_H \, P(p_H|{\rm data})$.  Interestingly, in this case, the result is not the maximum-likelihood solution $H/N$, but rather $(H+1)/(N+2)$, a ``correction'' originally proposed by Laplace.\footnote{If instead of a flat prior, one uses the Bures prior, one arrives at $(H+0.5)/(N+1)$.}  Note that instead of being undefined for $N=0$, this formula returns the mean of the (flat) prior distribution, $0.5$; and that it can never reach $0$ or $1$.   The subtlety here is again one of inversion.  Adjusting $p_H$ to minimize the $\chi^2$ (essentially, the mean squared-deviation of our observations from the model prediction) would yield the maximum-likelihood result (in the above example, $p_H$=0); this is not the same thing as minimizing the mean squared-error of our estimate for $p_H$ from its true value.

\subsection{A Quantum Example}
\label{IA1}

The relation of this formalism to quantum measurement theory arises because the quantum state (by which I mean either a wave vector $\ket{\psi}$ or a density matrix $\rho$) is fundamentally a way of making predictions about future observations: a description of our knowledge, or ``prior.''  When we gain information by performing a measurement, our knowledge of the system (and our best estimates of future probabilities) changes.  At the simplest level, this is the origin of ``collapse.''  Let us consider a simple example central to quantum optics.  Consider an atom initially in the state $\ket{+} \equiv \left(\ket{g}+\ket{e}\right)/\sqrt{2}$.  Imagine letting this atom evolve for one half-life, all the while observing it with a 100\%-efficient photodetector.  If you detect a photon, the atom must have been in $\ket{e}$, so you have that new information about the original state -- but of course, since it has decayed, it must be in $\ket{g}$: such a measurement both provides information about the initial condition and simultaneously modifies the state.  

On the other hand, if you wait one half-life and observe no photons -- what state should you conclude the atom is in?  Since it hasn't decayed, you might suspect that it is still in $\ket{+}$.  This can clearly be seen to be incorrect by going to extremes: if you waited for 100 years and still observed no photon, you would presumably reason that any excited atom would have decayed by now, so the atom must have been in the ground state all along.  Even though no photon was emitted, your description of the quantum state should change from $\ket{+}$ to $\ket{g}$.  {\it Observing nothing is also an observation.}  So how should we apply this intuition at intermediate times, such as after a single half-life?  It turns out that the Bayesian methods introduced above are perfectly adequate.  Our prior probabilities (given by the initial quantum state) are $P(g)=P(e)=1/2$, and if we consider ``g'' and ``e'' to be two different models of the atom and a single photon ``$\gamma$'' to be the observation, we can write our conditional probabilities $P(\gamma | e) = 1/2$ and $P(\gamma | g) = 0$.  The calculation is straightforward:
\begin{eqnarray}
P(g|\gamma) \propto & P(g) P(\gamma | g) & = \frac{1}{2}\cdot 0 \; \nonumber \\
P(e|\gamma) \propto & P(e) P(\gamma | e) & = \frac{1}{2}\cdot\frac{1}{2} \; ,
\end{eqnarray}
from which we see that $P(g|\gamma)=0$ as expected and $P(e|\gamma)=1$.  Similarly,
\begin{eqnarray}
P(g|{\rm no }\;\gamma) \propto & P(g) P({\rm no }\;\gamma | g) & = \frac{1}{2}\cdot 1 \; \nonumber \\
P(e|{\rm no }\;\gamma) \propto & P(e) P({\rm no }\;\gamma | e) & = \frac{1}{2}\cdot\frac{1}{2} \; ,
\end{eqnarray}
from which we conclude that after {\it no photon} is observed, the probability of $g$ is twice that of $e$, and hence must
be $2/3$.  Indeed, the result of this non-observation of a photon is to ``collapse'' the state to a new state $\sqrt{2/3}\ket{g} + \sqrt{1/3}\ket{e}$ as suggested by this classical analysis (though of course the Bayesian reasoning
does not tell us anything about the phase relationship between $g$ and $e$; to understand that it is conserved, we will need the full quantum formalism, to be discussed in section \ref{POVMs}).  

This example can also be seen as a simple case of a more general observation: {\it not all measurements are projections}.  By observing no photon to be emitted for some length of time, we have acquired some information about the state, but a limited amount.  This modifies the state vector, but it does not project it onto one of two alternative
eigenstates; in fact, the final states which result from the two possible measurement outcomes are not even orthogonal to one another.

\section{Projective measurement, density matrices, and decoherence}
\label{IB}

\subsection{Review of projective measurement}
\label{IB0}
Before treating the modern approaches to measurement which can properly describe situations such as the ``non-observation'' we just discussed, we should briefly review the ``standard'' measurement postulate treated in most of our textbooks, also referred to as ``von Neumann'' or ``projective'' measurement.  By this postulate, objects which can be measured are ``observables'' corresponding to Hermitian operators.  Such an operator $\hat{A}$ has a set of orthonormal eigenstates $\ket{n}$ such that $\hat{A}\ket{n} = a_n \ket{n}$, where the (real) eigenvalues $a_n$ are the only possible results of a measurement of the observable corresponding to $\hat{A}$.  From the orthonormality relation $\inner{m}{n}=\delta_{mn}$ and the completeness relation $\sum_m\proj{m}=I$, one can show that any state $\ket{\psi}$ may be decomposed in this eigenbasis, as $\ket{\psi}=\sum_ic_i\ket{i}$, with the coefficients $c_i$ given by the inner product $\inner{i}{\psi}$.  The first part of the measurement postulate is that when $\hat{A}$ is measured, a result $a_i$ will be found with probability $P_i = |c_i|^2 = \left|\inner{i}{\psi}\right|^2$.  The second part of the measurement postulate is based on the requirement that measurements be repeatable, essentially related to the idea of updating our probability distribution based on our observations.  If we have already measured $\hat{A}$ to be $a_i$, then we expect that a second measurement will also yield $a_i$, with 100\% probability.  This requires that all the coefficients $c_{j\neq i}$ vanish: the state thus ``collapses'' into the corresponding eigenstate $\ket{i}$.

This statement neglects the possibility of degenerate eigenvalues (with corresponding eigen{\it spaces} rather than eigenstates).  This is of particular importance once we wish to think about multipartite systems; since an observable of one subsystem does not depend directly on the state of another subsystem, it is automatically degenerate.  So what is the full state of the whole system after a measurement?  Consider, for example, an entangled state.  For the sake of argument, imagine the coherent superposition of my holding a black chess pawn in my left hand and a white pawn in my right, or the reverse: $\left(\ket{B_L}\ket{W_R}+\ket{W_L}\ket{B_R}\right)/\sqrt{2}$.  If I open my right hand, the probability of finding a white pawn there is $p(W_R)=1/2$.  Of course, if I find this white pawn, the probability of finding a black pawn there drops instantly from $1/2$ to $0$; this is the classical ``collapse'' of probabilities.  But it is equally clear that the probability of finding a white pawn in my {\it left} hand immediately jumps to $1$, even though I have only made a measurement on the right hand.  (Note that such a ``nonlocal'' collapse is entirely classical, and arises because of correlations we can understand through a common cause; the nonlocality inherent in quantum entanglement is of course deeper.)  Classically, the analysis is simple; there are four possible models: $B_LB_R, B_LW_R, W_LB_R, {\rm and }W_LW_R$.  When I find a white pawn in my right hand, the 2nd and 4th models have likelihoods of 1, while the 1st and 3rd have likelihoods of 0.  I thus update my prior by multiplying the individual model probabilities by these likelihoods and renormalizing.  Since my prior state was a 50/50 superposition of models 2 and 3, only model 2 ($B_LW_R$) survives.  The quantum-mechanical projection postulate is the natural extension of this classical idea: each {\it subspace} corresponding to a particular outcome gets scaled up in amplitude according to its likelihood.  Of course, if the event in question is observation of an eigenvalue $a_i$, the likelihoods are 1 for any eigenstate with this eigenvalue and 0 for any other eigenstate.  We therefore {\it project} onto the subspace with eigenvalue $a_i$, and renormalise:
\begin{equation}
\label{update1}
\ket{\psi_f} = \frac{{\rm Proj}(i) \, \ket{\psi_i}}{\sqrt{\bra{\psi_i}\,{\rm Proj}(i)\,\ket{\psi_i}}} \; .
\end{equation}

\subsection{Density matrices}
\label{IB1}

We introduce density matrices as a more general description of a quantum state, because not all states of knowledge can be described by wave vectors.  Specifically, consider again the behaviour of an atom which can spontaneously emit.  Let us begin with an atom in the excited state and 0 photons, writing this state as $\ket{e,0\gamma}$.  If we imagine, as before, that we have a collection system able to capture 100\% of the emitted photons, then we can describe the state of the composite system after one half life by
\begin{equation}
\ket{e,0\gamma} \rightarrow \frac{\ket{e,0\gamma}}{\sqrt{2}} + \frac{\ket{g,1\gamma}}{\sqrt{2}}  \; .
\end{equation}
By detecting 0 or 1 photons, we could ``collapse'' the atomic state into e or g, respectively.  But in the more realistic case where the hypothetical photon flies off to infinity with no possibility of us observing it again, how shall we describe the resulting situation?  There is a 50/50 probability that the atom is in g or e, {\it certes}.  But what phase should we choose?  $\left(\ket{g}+\ket{e}\right)/\sqrt{2}$?  $\left(\ket{g}-\ket{e}\right)/\sqrt{2}$?  Any such superposition would possess some dipole moment\footnote{if not at $t=0$ then anyway after some time evolution}; what would break the symmetry and determine whether this dipole moment should be positive or negative (point to the right or the left)?  Clearly, spontaneous emission should not define any preferential direction in space, so we need a way to describe a 50/50 {\it mixture} of ground and excited states which nevertheless has a vanishing expectation value of the dipole moment (which is to say, of the operator $\ket{g}\bra{e}$ and its Hermitian adjoint).

Such {\it probabilistic} mixtures are called ``mixed states,'' to distinguish them from ``pure states,'' which are any states that can be written as state vectors.  Note that the concept of purity is basis-independent -- thus {\it superposition states} are still pure.  We seek a mathematical description of situations where, beyond the intrinsic quantum-mechanical uncertainties, we have some ignorance about the state of the system.  There is a chance $P_A$ that the system is in pure state $A$ and a chance $P_B$ that the system is in pure state $B$, for instance (where $A$ and $B$ could be the ground and excited states, as above).  In such a case, the rule for an expectation value is clear; the expectation value for any observable in the mixed state should be the average, weighted by the probabilities $P_A$ and $P_B$, of its expectation values in $A$ and $B$ individually.  The same weighted-average rule should hold for the probability of any given event (the probability of finding a given state being, after all, the expectation value of the projector onto that state).  This is the classical law of total probability: $P(x) = P(A)P(x|A) + P(B)P(x|B) + \ldots$, where $\{A,B,\ldots\}$ are any exhaustive set of mutually exclusive possibilities.  The problem drawing a direct quantum analog to this expression is that our description of the state is not a direct list of probabilities $P(x|A)$ but rather a state vector (or wave function, or list of probability {\it amplitudes}, if you prefer).  In particular, since $P(x)=|\psi(x)|^2$ is not linear, we cannot take $\psi(x)=\psi_A(x)P(A)+\psi_B(x)P(b)$, nor even $\psi(x)=\psi_A(x)\sqrt{P(A)}+\psi_B(x)\sqrt{P(b)}$; any such expression would supplement the desired (``classical'') weighted-average law with cross-terms.  The situation would be simpler if instead of describing a physical situation with a state function which appears quadratically in expectation values $\bra{\psi}X\ket{\psi}$ and in probabilities 
\begin{equation}
\label{inner}
P(i) =\left|\inner{i}{\psi}\right|^2 = \inner{i}{\psi}\inner{\psi}{i}
\end{equation}
we could identify a mathematical object which uniquely and completely described the situation, while appearing {\it linearly} in such expressions.  And this object leaps out at us from the right-hand side of Eq.~\ref{inner}.  The {\it projector} $\proj{\psi}$ has a one-to-one correspondence with the state vector $\psi$ and hence describes the physical situation equally well, but it appears linearly in the expression for probability.  It follows that if for a pure state we {\it define} the ``density matrix'' $\rho_{\rm pure} \equiv \proj{\psi}$, we can use the same rule $P(i) \equiv \bra{i}\rho\ket{i}$ for mixed states, simply by defining $\rho_{\rm mixed} \equiv \sum_m P_m \rho_m$, where $P_m$ are the probabilities to be in pure states described by the $\rho_m$, and the sum over $m$ is over any number of states which may be mixed, with no requirement that this set be complete, orthonormal, or anything else.  This construction allows us to use a single representation for any state -- pure or mixed -- and a single formula for probabilities or expectation values, the linearity of these formulas ensuring that the proper description of a mixed state is indeed the probability-weighted average of the pure states entering into the mixture.  It is important to note that this decomposition is not unique -- while $\rho$ is a Hermitian operator and therefore possesses a unique decomposition into an eigenbasis (aside from degeneracy), it will in general possess an infinite number of indistinguishable expansions of this more general form.  I will not discuss the mathematical properties of density matrices at more length, as they can be found in standard textbooks.

\subsection{Update rule for density matrices}
\label{IB2}
Having seen Eq.~\ref{update1} for the update rule we apply to a state vector upon obtaining a measurement result, we now need an analogous expression to update our information when it is written as a density matrix.  To do this, recall that the density matrix is a weighted average of projectors onto the different pure states the system ``might have been in.''  Upon observation of a measurement result $\ket{j}$, each one of these component pure states will simply be projected onto $\ket{j}$ (to within normalisation), such that
\begin{eqnarray}
& & \sum_m P_m \proj{\psi_m} \nonumber \\
& \Rightarrow & \frac{1}{N} \sum_m P_m \ket{j}\inner{j}{\psi_m}\inner{\psi_m}{j}\bra{j} \\
& = & \frac{1}{N} \sum_m \left[P_m P(j|m)\right] \proj{j} \; , \nonumber
\end{eqnarray}
where $N$ is a normalisation constant.  The expression on the right is straightforward to interpret in terms of Bayesian probabilities: the expression in square brackets is (again, modulo normalisation) the updated weighting probability $P(m|j)$; and the projector onto $\ket{j}$ appears because each component $\ket{\psi_m}$ collapsed onto that state when the measurement occurred.  The constant $N$ must $=\sum_m PmP(j|m)$, which is of course just $P_j$.  This expression can easily be generalised to the situation where $\ket{j}$ is replaced by a subspace, by replacing $\proj{j}$ with a projector ${\rm Proj}(j)$ which need not be rank 1:
\begin{equation}
\label{update2}
\rho_f = \frac{{\rm Proj}(j)\rho{\rm Proj}(j)}{P_j} \; .
\end{equation}

\subsection{Losing information}
\label{IB3}
Now we are ready to address the important question of lost information.  Unitary (Schr\"odinger) evolution conserves information -- that is to say, the entropy of a state remains constant, and the overlap of any two states also remains constant.  Irreversible loss of information only occurs for ``open systems,'' as is the case for classical irreversibility as well.  More rigorously, what this means is that when one subsystem is ``discarded,'' the evolution of the retained subsystem may be irreversible.  This is what happened in the example of section \ref{IB1}, of spontaneous emission.  When we hypothesized that a photon might be lost to infinity and never retrieved, the consequent behaviour of the atom was not unitary, and this necessitated the invention of the density-matrix formalism.  

Let us for convenience split the universe into two parts or subsystems, one which we are interested in studying and will term the ``system''; and everything else (beyond our ability to measure), which we will term the ``environment.''  Then we shall replace the basis states $\ket{i}$ by system-environment product states $\ket{i}_{\rm sys}\ket{j}_{\rm env}$.  (In the earlier example, $\ket{i}_{\rm sys}$ might be the ground and excited states of the atom while $\ket{j}_{\rm env}$ could be the $0$- and $1$-photon states of the field.)  Now for some state $\rho$ of the combined system and environment, let us calculate the expectation value of an operator $A$.    This is done by calculating ${\rm Tr} \rho A$, which involves summing all the diagonal elements of $\rho A$ in any complete basis, e.g., $\ket{i}\ket{j}$:
\begin{eqnarray}
\expect{A} \equiv {\rm Tr} \rho A & = & \sum_i \sum_j \bra{i_{\rm sys}}\bra{j_{\rm env}}\rho A \ket{i_{\rm sys}}\ket{j_{\rm env}} \nonumber \\
& = & \sum_i \bra{i_{\rm sys}}\left\{\sum_j \bra{j_{\rm env}}\rho A\ket{j_{\rm env}}\right\}\ket{i_{\rm sys}} \; .
\end{eqnarray}
Now, we have assumed that we can only carry out measurements on the {\it system}, and not on the environment.  Thus $A$ acts only on the system part of the state, and commutes with $\ket{j_{\rm env}}$:
\begin{eqnarray}
\expect{A} &=& \sum_i \bra{i_{\rm sys}}\left\{\sum_j \bra{j_{\rm env}}\rho\ket{j_{\rm env}}\right\} A \ket{i_{\rm sys}} \nonumber \\
& = & {\rm Tr}_{\rm sys} \rho_{\rm red} A \; , {\rm where} \\
\rho_{\rm red} & \equiv & {\rm Tr}_{\rm env} \rho \nonumber 
\end{eqnarray}
is termed the ``reduced density matrix,'' and represents the density matrix for the system, once the environment is ``discarded'' (or ``traced over,'' in the mathematical jargon).  As can be seen clearly in the above sums, the partial traces ${\rm Tr}_{\rm sys}$ and ${\rm Tr}_{\rm env}$ are computed by taking the diagonal matrix elements between basis vectors for the system and environment individually.    Physically, tracing over the environment reflects ignoring the state of the environment.  One way to conceive of doing this is to imagine an observer measuring the environment in some basis ($\ket{k}$).  For any outcome he obtains, the system will be left in a density matrix $\bra{k}\rho\ket{k}$ (with probability given by the trace of that -- unnormalised -- density matrix, as I leave as an exercise).  If we have no access to this observer's result, our system will be nonetheless left in a weighted average of all possible $\bra{k}\rho\ket{k}$'s.  Summing these unnormalised matrices over a complete set of $k$ includes this weighting automatically -- and since this sum is a trace (${\rm Tr}_{\rm env}$), it is basis-independent: as one should perhaps have expected, the (average, mixed) state of our system cannot depend on which basis some unknown observer chooses to measure the environment in.

\begin{figure}[tb]
\centering
\includegraphics[width=4in]{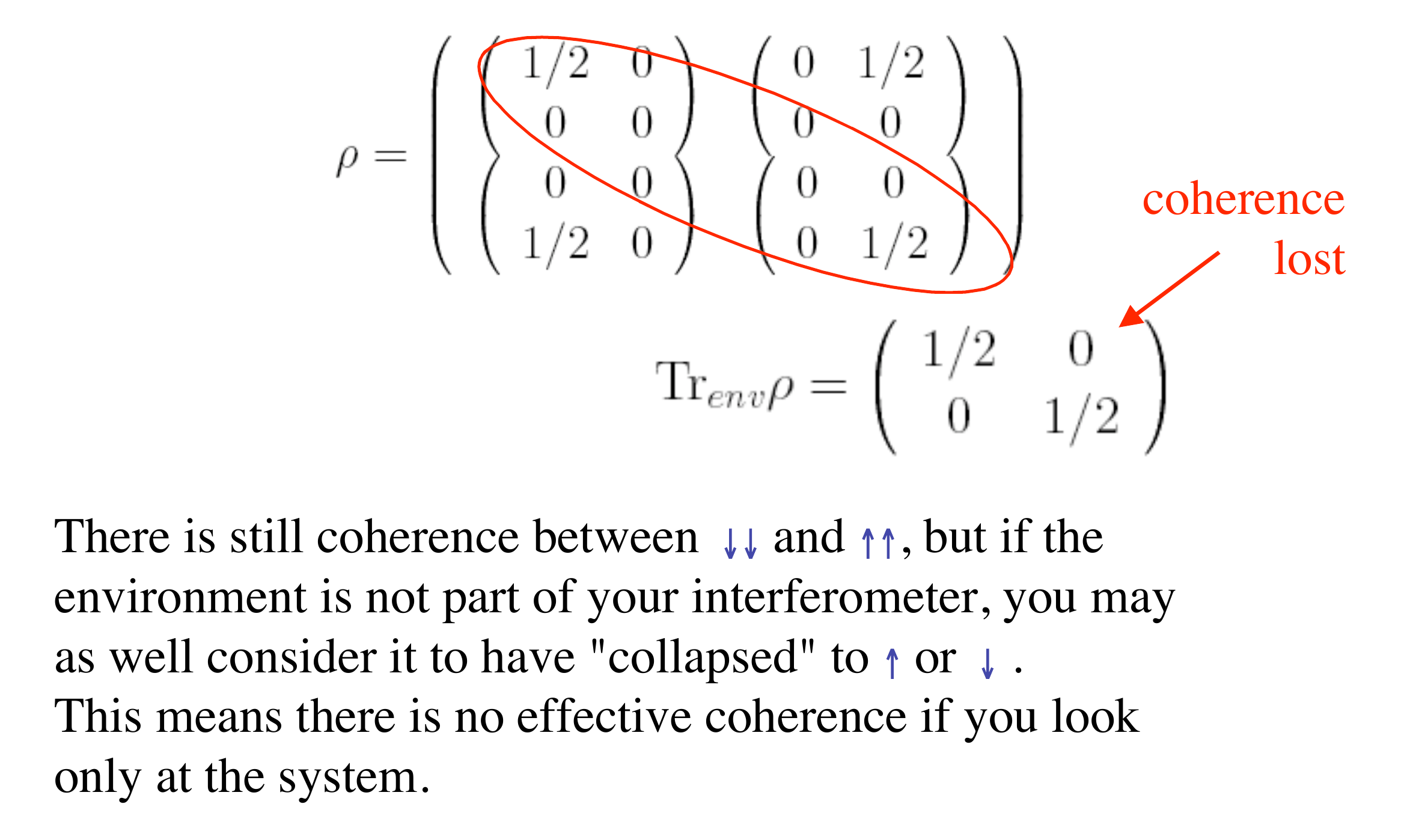}
\caption{The effects on the 4x4 density matrix for atom and field of tracing over the unobserved field state}
\label{matrix}
\end{figure}

Mathematically, taking this trace over the environment means that we retain only terms diagonal in the environment variables -- no cross-terms (coherences) are retained between terms corresponding to different states of the environment.  This is illustrated in Figure \ref{matrix} for the case of the spontaneously decaying atom in state $\left(\ket{e0}+\ket{g1}\right)/\sqrt{2}$.

The full state of the entangled atom and field is given by this 4x4 matrix.  The state of the system when no photon is observed is given by the upper-left 2x2 block, while the state when 1 photon is observed is given by the lower-right 2x2 block.  The final state of the system if we do not know which occurred is given by

\begin{eqnarray}
\rho_{\rm red}  =  \bra{0_{\rm env}}\rho\ket{0_{\rm env}} \; & + & \;  \bra{1_{\rm env}}\rho\ket{1_{\rm env}}  \nonumber \\
 =  \left( 
\begin{array}{cc}
1/2 & 0 \\
0 & 0 \\
\end{array} \right)  & + & \left(
\begin{array}{cc}
0 & 0 \\
0 & 1/2 \\
\end{array} \right)
\; \;
=  \left(
\begin{array}{cc}
1/2 & 0 \\
0 & 1/2 \\
\end{array} \right) \; . 
\end{eqnarray}

This expression is at the heart of the effective decoherence which is characteristic of open systems, but also of measurement; there is, of course, no difference between an ``environment'' treated as above (which acquires information about the state of the system through an entangling interaction) and a measuring device, save for intent.  As Feynman's Rules for interference teach us, if there is any way -- even in principle -- to tell which of two histories (Feynman paths, if you like) was followed, then no interference may occur between these paths: their coherence is lost (see section \ref{comp} for further discussion of this).  Clearly, by observing the environment, one could determine whether the atom was in $\ket{g}$ or $\ket{e}$; this is why the reduced density matrix has $0$'s on the off-diagonals, representing vanishing coherence between the two.  There can be no dipole moment, since a nonzero dipole arises from interference between states of different parity.  Naturally, the full system still has perfect coherence between $\ket{g1}$ and $\ket{e0}$; but this coherence can only be studied by manipulating the two systems together; once we limit ourselves to studying a subsystem (or once the environment has such a large number of degrees of freedom that, as in statistical thermodynamics, it is inconceivable in practice that it could be manipulated sufficiently coherently), this coherence becomes practically unobservable.  This is true, incidentally, even if there is no additional ``collapse'' process, and the evolution of the universe is described by the Schr\"odinger equation alone\ldots  In sum, effective ``collapse'' (really decoherence: the impossibility of observing any effects which depend simultaneously on the amplitudes for Schr\"odinger's cat to have been alive and to have been dead) arises in two steps: (1) the entanglement of two systems, and (2) the discarding (or neglecting) of one of these systems.

Importantly, the trace is basis-independent.  It is not necessary that some demon in the environment choose to measure whether there was 1 photon or 0, and acquire this information; the mere fact that this is {\it conceivable} guarantees that there is no interference.  If such a demon measured the field to be in the state $\left(\ket{0}+\ket{1}\right)/\sqrt{2}$, he would in fact collapse the atom into a state with a positive dipole moment.  This is the content of the ``quantum eraser'' idea.  However, if the demon attempted such a measurement, he would be equally likely to find  $\left(\ket{0}-\ket{1}\right)/\sqrt{2}$, collapsing the atom into a state with negative dipole moment; if we {\it ignore} the demon's subsystem, the density matrix is the same regardless of what basis he picks for a measurement.  This is a reflection of the fact that no action that is performed on one subsystem can affect the statistics of another subsystem, once the two are no longer directly interacting.  This is one way to see that EPR correlations satisfy a ``no-signalling'' theorem, and do not violate relativistic causality.

\section{Generalized Measurement (POVMs)}
\label{POVMs}

\begin{figure}[tb]
\centering
\includegraphics[width=7in]{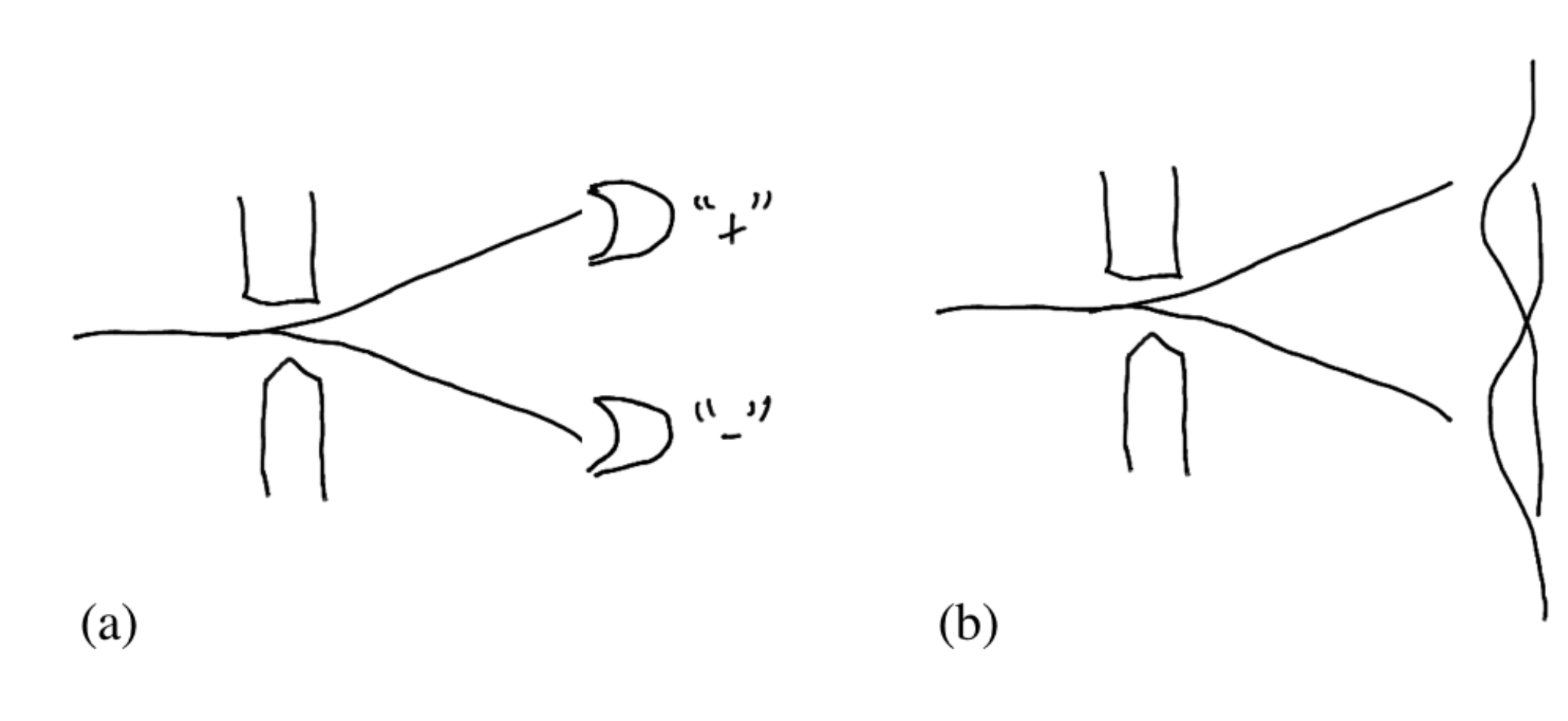}
\caption{A Stern-Gerlach measurement.  a. The idealized ``two-dimensional'' picture of a projective measurement.  b. The reality: infinite-dimensional, and with finite uncertainties.}
\label{SG}
\end{figure}

All measurement is indirect.  By this I wish to say that when we talk about studying a system $A$, we really mean that the system interacts with a meter $B$, and we look at the meter.  Really, this chain typically has many more steps, culminating whenever you feel like truncating it -- when the information is written on your hard drive, or displayed on your computer screen, or when your eyes absorb the photons from the screen, or after your brain processes the information?  But to understand the effect on the system and the maximum amount of information extractable, it is enough to consider the interaction of $A$ with $B$.  Now, the traditional (von Neumann) view of measurement is in terms of projectors onto complete sets of basis states for Hermitian observables; obviously, when in the real world we carry out a Stern-Gerlach experiment and a silver atom causes a spot to appear on a screen, this spot does not live in a two-dimensional Hilbert space and spit out a ``$+\hbar/2$'' or a ``$-\hbar/2$'' (Figure \ref{SG}).  Only given some knowledge of the system do we make the {\it approximation} that spots in certain regions {\it nearly} guarantee one eigenstate and spots in others nearly guarantee the other, recognizing that some experimental uncertainty is always unavoidable.  Any real observation is thus viewed as an approximation to the Platonic ideal of a measurement.  A more appropriate view, I would maintain, is that any correlation can provide information, and thus constitutes a measurement.  Projection operators are merely one idealization of this idea -- and interestingly, they are not even always the optimal strategy, depending on one's goal.  This is the motivation for discussing ``generalized measurement,'' a formalism which is intended to encompass both von Neumann-style measurement and the wide range of real-world techniques for extracting information from a system.

Let us explicitly consider the interaction between a system of interest and a second, ``meter'' system.  To accomplish an ideal von Neumann measurement, we would like an interaction to create perfect entanglement between the system and the meter, which should have the same dimensionality as the system.  For instance, if we wish to observe whether an atom is in $\ket{g}$ or $\ket{e}$, we need some 2-level measuring device, such as a spin which can be in state $\ket{\downarrow}$ or $\ket{\uparrow}$.  Initially, the atom is in an unknown state, and the spin is ``initialized,'' e.g., to $\ket{\downarrow}$.  Our goal would be to design an interaction which would lead to the evolution
\begin{eqnarray}
\ket{g}\ket{\downarrow} & \rightarrow & \ket{g}\ket{\downarrow} \nonumber \\
\ket{e}\ket{\downarrow} & \rightarrow & \ket{e}\ket{\uparrow} \; ,
\end{eqnarray}
for instance, so that later observation of the spin would inform us as to the state of the atom.  (The astute reader with some familiarity with quantum information will have observed that this interaction is essentially a controlled-NOT.)  Meanwhile, since the spin has complete information about the state of the atom, this measurement has destroyed all coherence between $e$ and $g$.

But recall our earlier discussion of an ``incomplete'' measurement in section \ref{IA1}, where only giving the atom one half-life to interact with the field failed to maximally entangle the two systems:
\begin{eqnarray}
\ket{g}\ket{0} & \rightarrow & \ket{g}\ket{0} \nonumber \\
\ket{e}\ket{0} & \rightarrow & \frac{\ket{e}\ket{0}+\ket{g}\ket{1}}{\sqrt{2}} \; .
\end{eqnarray}
Here, observing the field to be in ``0'' or ``1'' gives some information about the state of the atom, but not complete information; it is a therefore a reasonable model for at least one class of real-world measurements with finite uncertainty.  Another example of a measurement with finite uncertainty is shown in Fig.~\ref{SG}b.  In the Stern-Gerlach effect, the (two-dimensional) spin of the atom is coupled to the (continuous) momentum of the atom, which is what is eventually read off (ideally), in the far field.  In this case, the infinite dimensionality of the momentum ``pointer'' is merely a technical detail, perhaps a weakness, but any true description of practical measurements should take into account the fact that the number of possible outcomes may be larger than the dimensionality of the system Hilbert space alone -- perhaps even infinite.  In the present situation, this may be construed as a minor technical nuisance, but as we shall see later, there are cases in which there is an actual advantage to performing such measurements.

In the case of projective measurements, the probability of outcome $i$ was given by the expectation value of the projector onto $i$: $P_i = \expect{{\rm Proj}(i)} = {\rm Tr}(\rho {\rm Proj}(i))$.  If the $i$'s come from a complete set of orthonormal states, then the completeness relation $\sum_i {\rm Proj}(i) = I$ ensures that $\sum_i P_i = 1$, as desired.  For a more general measurement, we still expect a basis-independent expression linear in $\rho$ to describe the probability of a given outcome, and a small generalisation suffices:
\begin{equation}
P_i = {\rm Tr}(E_i\rho) \; ,
\end{equation}
where the $E_i$ need not be projectors, but must be {\it positive} (to guarantee that $P_i$ is always positive) and must sum to the identity: $\sum_i E_i = I$.  These operators can be written in terms of ``measurement operators'' $M_i$ as follows:
\begin{equation}
E_i = M_i^\dagger M_i \; .
\end{equation}
Note that while the $M_i$ determine the $E_i$, the reverse is not true.  Any $(M_i^\dagger U^\dagger) (U M_i) = M_i^\dagger M_i$, after all.  What is the meaning of the measurement operators?  If an outcome $i$ is found, then $M_i$ describes the effect on the state: $\ket{\psi} \rightarrow M_i\ket{\psi}$.  The simplest example is given by $M_i = \proj{i}$, which leads automatically to $E_i = \proj{i}$, and recovers the usual projective measurements, including both the probability formula and the projection postulate.  But consider the case of spontaneous emission once more: the probability of detecting a photon depends on the initial probability of being in the excited state (multiplied by some ``efficiency'' $\eta$), but upon detection of this photon, the atom is left in the ground state.  The measurement operator for such a decay observation is $M_1 = \sqrt{\eta} \ket{g}\bra{e}$.  (This is stated here without proof, but it can be derived directly from the form of the interaction Hamiltonian which coupled the atom to the field, creating a photon through the term $a^\dagger \ket{g}\bra{e}$.)  It is easy to see that $E_1 = M_1^\dagger M_1 = \eta \proj{e}$, so that the probability of detecting a photon is given by the probability for the atom to be in $e$, multipled by $\eta$.  The corresponding operator for the detection of 0 photons must be $E_0 = I - E_1$ since there are only two possible outcomes in this example: 
\begin{equation}
E_0 = I-E_1 = \proj{g} + (1-\eta) \proj{e} \; ;
\end{equation}
The probability of detecting no photon is the sum of two terms, one for an atom in the ground state and the other for an atom in the excited state which failed to emit a photon.    As we saw before, such a non-detection event makes it more likely -- but not certain -- that the atom is in $g$.  How shall we see this in the update rule?  Again, there are multiple ``square roots'' of $E_1$, and the physically relevant one depends on the details of the interaction Hamiltonian, but the simplest case is easily seen to be
\begin{equation}
M_0 = \proj{g} + \sqrt{1-\eta}\proj{e} \; .
\end{equation}
An atom initially in $c_g\ket{g} +c_e\ket{e}$ is left in a state $c_g\ket{g} + \sqrt{1-\eta}c_e\ket{e}$ upon non-observation of a photon.  In the example we previously considered of $c_g=c_e=1/\sqrt{2}$ and $\eta=1/2$, 
\begin{equation}
M_0\ket{\psi} = \frac{\ket{g}+\frac{1}{\sqrt{2}}\ket{e}}{\sqrt{2}} \; ,
\label{Mpsi}
\end{equation}
confirming our classical conclusion that $e$ is half as likely as $g$, implying a $33\%$ probability to be in $e$ and a $67\%$ probability to be in $g$.  Note that the state in Eq.~\ref{Mpsi} is not normalized; just as in the case of projection operators, the norm-squared of this resultant state ($\bra{\psi}M_i^\dagger M_i\ket{\psi}=\expect{E_I}$) gives the probability of the result -- here $50\%$ (for an atom starting in g) plus $25\%$ (for an atom starting in e but failing to decay) for a total of $75\%$.  But when that event occurs, we are left in a reweighted {\it coherent} superposition of $e$ and $g$.  The reweightings of the probabilities can be seen to be simply $\bra{g}M_i^\dagger M_i\ket{g}$ and $\bra{e}M_i^\dagger M_i\ket{e}$, which are nothing but the Bayesian likelihoods of the models $g$ and $e$, respectively (the conditional probability of outcome $i$ {\it given} an initial state of $g$ or $e$).

The update rule written for pure states can easily be rewritten in terms of density matrices, and proves again to be linear, which guarantees that the following update rule is generally applicable to any $\rho$, pure or mixed:
\begin{equation}
\rho \rightarrow \frac{M_i \rho M_i^\dagger}{P_i} \; ,
\end{equation}
where $P_i$ is still given by $\expect{E_i} = {\rm Tr} \rho M_i^\dagger M_i$.  Again, it is easy to verify that this reduces to the update rule for projective measurements when the $M_i$'s are taken to be a complete set of orthonormal projection operators.  These generalized measurements are also often referred to as ``POVMs,'' for ``positive operator-valued measurements,'' although this is such an uninformative mouthful that some authors have been known to resist even including it in their publications, arguing that ``POVM'' should by now stand on its own as a synonym for generalized measurement.  Naimark's theorem shows that {\it any} POVM can be accomplished in the manner I have been outlining here, that is, by coupling the system of interest to some (potentially higher-dimensional) pointer system, and then making a projective measurement on the pointer.  

Quantum optics students should be familiar with the treatment of spontaneous emission on the Bloch sphere, where the inversion decays with a timescale $T_1$ while the coherences decay with a timescale $T_2$, where for pure radiative decay with no broadening, $T_2 = 2T_1$ (which can be understood at one level by arguing that if amplitudes decay as $e^{-t/T_2}$, then the excited state probability, being proportional to the square of an amplitude, will decay as $e^{-2t/T_2} \equiv e^{-t/T_1}$).  This means that as an atom initially in a pure state on the surface of the Bloch sphere decays, it does not follow a straight trajectory towards the South pole ($g$), but rather curves around, outside this na\"ive path.  From the perspective of generalized measurements, it is easy to derive this result, which most of us were taught to put in by hand when learning to write down the optical Bloch equations.  Either a photon was emitted or not; if we do not know the result of this hypothetical measurement, then we must ``trace over'' all outcomes -- here, this simply means summing over all $M_i$ with the appropriate weightings:
\begin{equation}
\rho \rightarrow \sum_i P_i \frac{M_i \rho M_i^\dagger}{P_i} = \sum_i M_i\rho M_i^\dagger \; .
\end{equation}
Since this describes decoherence (the evolution from a pure state into a mixed state), there is no corresponding form in terms of state vectors.  

\subsection{An example: unambiguous state discrimination}
\label{USD}

A prototypical case in which one can see that coupling to a higher-dimensional system may actually be advantageous is the problem of unambiguous state discrimination.  The problem is as follows: suppose that you are provided with a single quantum system, which is guaranteed to be prepared in either state $\ket{a}$ or state $\ket{b}$.  How well can you tell which of the two states you have?  Clearly, if $\inner{a}{b}=0$, a single projective measurement gives you the answer with certainty.  But if the overlap does not vanish, it is not possible to unambiguously discriminate the two states all of the time.  For simplicity, suppose the two states lie in a 2-dimensional Hilbert space; they could, for instance, refer to a horizontally-polarized photon and a $45^\circ$-polarized photon.   No projection which is guaranteed to detect every H can reject every 45.  At least two different goals are conceivable; one might wish to simply minimize the error rate (``minimum-error discrimination''), or one might wish to answer {\it with certainty} as often as possible (``unambiguous state discrimination'').  The former case was solved by Helstrom, and the solution is more or less what one might expect: one chooses a basis symmetrically positioned about $\ket{a}$ and $\ket{b}$ -- in our 0/45 example, one would project onto $-22.5^\circ$ or $+67.5^\circ$.  It is easy to see that (for instance) a $+67.5^\circ$ detector is much more likely to fire if the photon is polarized at $45^\circ$ than if it is polarized at $0^\circ$ (roughly 5.8 times so); this leads to an error rate of $\frac{1}{2}\left(1-\sqrt{1-\left|\inner{a}{b}\right|^2}\right)$ if the two possibilities were equally likely to begin with: about $15\%$ in our simple example.  

But how could one ever be {\it certain} that the state was $\ket{a}$?  The only way is to be certain that it is {\it not} $\ket{b}$, which suggests performing a projective measurement onto the orthogonal state $\ket{\bar{b}}$.  Unfortunately, to be certain that the state {\it is} $\ket{b}$, one would have to project onto $\ket{\bar{a}}$, which is of course not orthogonal to $\ket{\bar{b}}$.  Thus on any individual case, one must choose to do one or the other.  For instance, we could project onto the 0/90 (H/V) basis for the photon, but when an H is observed, we have no way of being certain whether the photon was polarized along H or along 45, and must simply report `Don't know.'  Only when we observe a V can we conclude that the photon {\it could not} have been H, and thus must have been alone $45^\circ$.  Since only half the photons were presumed to be along $45^\circ$, and only half of these are transmitted by a V polarizer, this strategy succeeds only 1/4 of the time, or more generally $\frac{1}{2}\left(1-\left|\inner{a}{b}\right|^2\right)$.  Interestingly, information-theoretic arguments show that the optimum success rate is $1-\left|\inner{a}{b}\right|$, or about $29\%$ for our example.  There is no way to achieve this optimum with projective measurements, and the reason can be understood in the following simple way.  We are studying a two-dimensional system, and wish to be able to report either `a' or `b' when we find the appropriate result -- but, aware that we will be unable to do so correctly $100\%$ of the time, we must also report `DK' (``Don't know'') on occasion.  In other words, we need to perform a measurement with 3 possible outcomes on this two-dimensional system, in violation of the assumptions of projective measurement.  We hence need a generalized measurement, with three ``POVM elements'' $E_i$, corresponding to the three results `a,' `b,' and `DK.'  As mentioned earlier, the strategy for doing this is to ``expand the Hilbert space,'' for instance by allowing the system to interact with a three-dimensional pointer system.  

\begin{figure}[tb]
\centering
\includegraphics[width=7in]{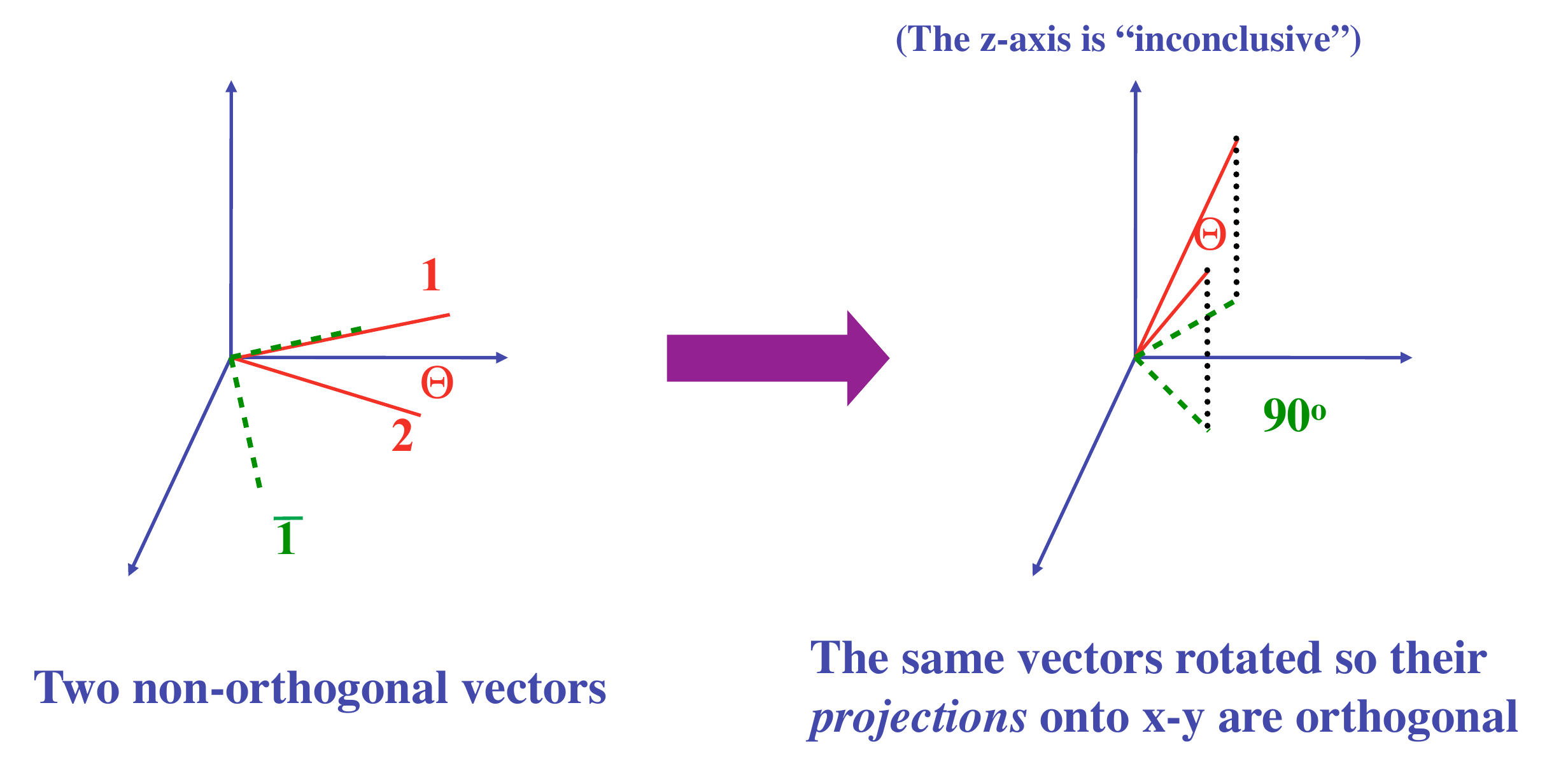}
\caption{The POVM for unambiguous discrimination of two non-orthogonal states in a 2D Hilbert space, viewed geometrically as a rotation into a third dimension.}
\label{3d}
\end{figure}

Another way to think about this is to recognize that one would like the two nonorthogonal states $\ket{a}$ and $\ket{b}$ to evolve into perfectly distinguishable (i.e., orthogonal) pointer states $\ket{A}$ and $\ket{B}$ which we can observe and announce.  However, since unitary evolution preserves the overlap, there is no purely unitary process which can accomplish this.  Something non-unitary must be done; one non-unitary operation is projection.  We can allow the system to interact with a pointer so that the overall state expands into a three-dimensional state, but then {\it project} onto the AB subspace, so that the two states are orthogonal {\it after projection}.  Mathematically, the unitary evolution could be written
\begin{eqnarray}
\ket{a} & \rightarrow & u\ket{A}+v\ket{DK} \nonumber \\
\ket{b} & \rightarrow & w\ket{B}+x\ket{DK} \; , 
\end{eqnarray}
where $A$, $B$, and $DK$ are three orthogonal pointer states indicating `we are sure the state is a,' `we are sure the state is b,' and `we don't know,' respectively.  Unitarity guarantees that $|vx| = \left|\inner{a}{b}\right|$, so the ``failure probability,'' that is, the probability of being required to admit we don't know, is
\begin{equation}
P_{DK} = \frac{|v|^2+|x|^2}{2} \geq |vx| = \left|\inner{a}{b}\right| \; .
\end{equation}
The maximum success rate, achieved when $v$ and $x$ are chosen to saturate the inequality, is as promised $1-\left|\inner{a}{b}\right|$, or about $29\%$ for the example we have discussed.  This transformation can be pictured as a rotation of the 2-dimensional vectors into a 3-dimensional space, as in Fig.~\ref{3d}, where the z-axis represents the $DK$ state, and the rotation angle is chosen to make the projections of $a$ and $b$ on the xy plane orthogonal (these projections are then the $A$ and $B$ vectors).  

Expressed in the POVM formalism, the operators $E_A$ and $E_B$ are not full projectors onto states $a$ and $b$, for these would not sum to unity.  Instead, three operators are required, such that $I = E_A + E_B + E_{DK}$.  In the optimal solution, $E_A = \kappa \proj{\bar{b}}$, $E_B = \kappa \proj{\bar{a}}$, and therefore $E_{DK} = I - \kappa \left(\proj{\bar{b}}+\proj{\bar{a}}\right)$.  The constant $\kappa$ must be chosen so that $E_{DK}$ remains a positive operator, and when it is maximized under this constraint, one achieves the best-case unambiguous discrimination.    \begin{figure}[tb]
\centering
\includegraphics[height=3in]{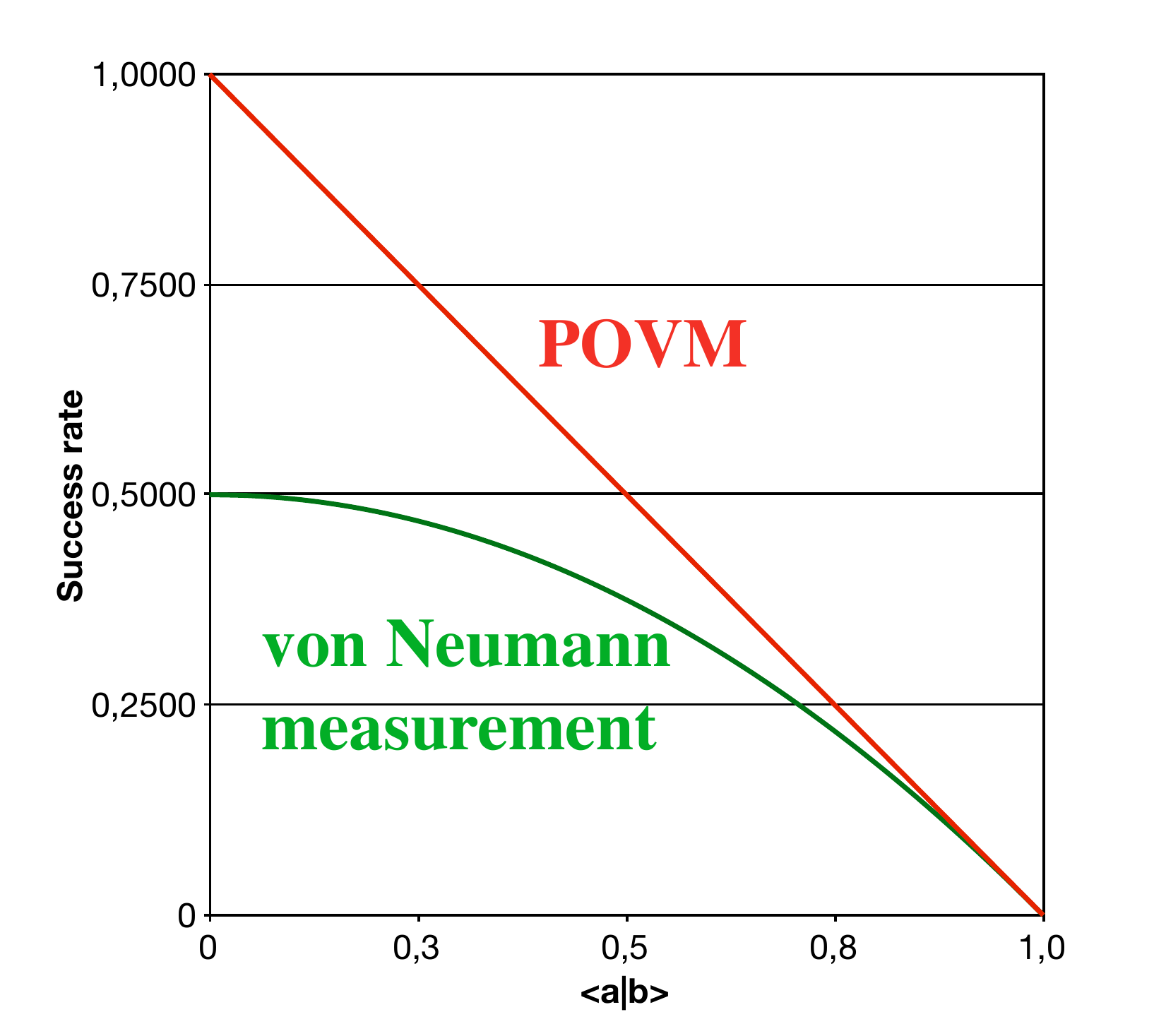}
\caption{The maximum success probabilities for unambiguous discrimination of two states, as a function of their overlap, for projective measurements and POVMs.}
\label{eff}
\end{figure}
Fig.~\ref{eff} shows how the success rate of this POVM-based solution compares with the maximum achievable using projective measurements.  Note that since no projective strategy can work for more than one state at a time, the success rate with projective measurements is strictly less than $50\%$ except when $a$ and $b$ are orthogonal; on the other hand, the POVM success rate grows smoothly to $100\%$.  In higher dimensions, the advantage is even greater, because in the most general case, one can distinguish up to $d$ states in a $d-$dimensional space (the states must be linearly independent for unambiguous discrimination to be possible), and there will usually be no projective measurement basis which can unambiguously identify more than a single one of these $d$ candidate states; an example is shown in Fib.~\ref{3DPOVM}.

\begin{figure}[tb]
\centering
\includegraphics[width=7in]{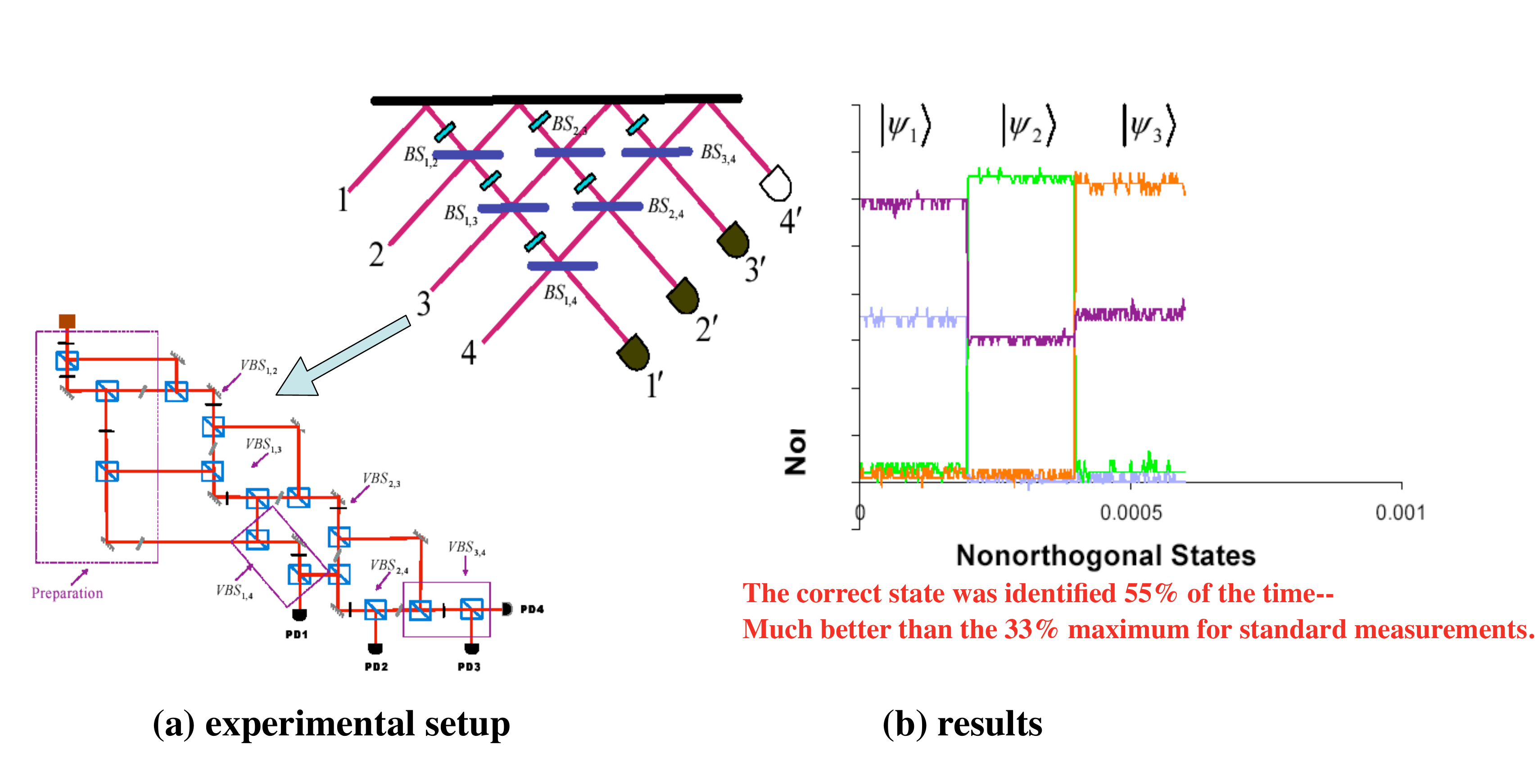}
\caption{In our group, we carried out an optical experiment to distinguish among three qutrit states, achieving a 55\% success rate (to be compared with the $<33\%$ maximum for projective measurements).}
\label{3DPOVM}
\end{figure}

\section{Complementarity: Feynman's Rules and the Quantum Eraser}
\label{comp}
We are all familiar with Bohr's complementarity principle, and the conclusion of the Bohr-Einstein debates, that measuring which-path information destroys interference.  In section \ref{IB3}, we saw how coupling to an environment and {\it ignoring} the final state of the environment accomplishes this, by erasing the off-diagonal terms in $\rho$.  The implication is that actually measuring ``which path'' a particle takes is not required: the mere possibility {\it in principle} of measuring this -- that is to say, the existence of anything in the environmental state from which such information could conceivably be drawn -- is already enough.  This is the content of ``Feynman's rules for interference'': 
\begin{quote}
I - if two or more fundamentally {\it indistinguishable} processes can lead to the same final event (e.g., a particle appearing at a given point on a screen), then add the complex amplitudes for these processes, and take the absolute square of the result to find the probability of the event.\\
II - if processes are distinguishable, even {\it in principle}, then take the absolute squares of their amplitudes individually, and add the resulting probabilities to find the total probability of the event.
\end{quote}
This language can still confuse people -- what exactly do we mean by ``in principle''?  For instance, as Einstein repeatedly pointed out, it is certainly possible to measure which slit a particle passes through.  The question is whether that possibility exists {\it after} the particle has reached the screen.  If the experiment was built in such a way that once the particle reaches the screen, there is {\it no longer} any conceivable way to measure which slit it followed, then interference occurs; otherwise, it does not.  A more modern perspective would be to say that one should calculate the probability of a given final state of the whole universe, not just of a single particle.  If two different processes can lead to the same final state of {\it everything}, then there is no way to tell from that final state which process occurred.  If, on the other hand, some demon has written down which slit the photon went through, then there are two different final states for the universe: one with a photon on the screen and the words ``upper slit'' recorded in the demon's lab book, and one with the same photon at the same point on the screen but the words ``lower slit'' written in the lab book.  

Let us see how this applies to our earlier description of measurement as based on an interaction of the system with a ``pointer'' or environment.  Let us, without much loss of generality, consider the state of the system in a two-path interferometer, which I will write $\psi_s = \psi_a + e^{i\phi}\psi_b$, where $\psi_a$ and $\psi_b$ are meant to represent the wave functions corresponding to particles traversing slits (or generalized ``paths'') $a$ and $b$, respectively.  The final probabilities (or intensity pattern) are given by 
\begin{equation}
|\psi_s|^2 = |\psi_a|^2 + |\psi_b|^2 + e^{i\phi}\psi_a^*\psi_b + e^{-i\phi}\psi_a\psi_b^* \; ,
\end{equation}
where the two cross-terms on the right are of course the interference terms. 
But now let the system interact with a measuring apparatus ``MA,'' such that $\psi_s \rightarrow \psi_s\ket{MA}$.   In particular, we imagine that
\begin{eqnarray}
\label{MA}
\psi_a & \rightarrow & \psi_a\ket{A} \; {\rm and} \nonumber \\
\psi_b & \rightarrow & \psi_b\ket{B} \; .
\end{eqnarray}
Repeating the calculation of the intensity pattern, we now have
\begin{equation}
\label{MA2}
|\psi_s|^2 \rightarrow |\psi_a|^2 \inner{A}{A} + |\psi_b|^2 \inner{B}{B} + e^{i\phi}\psi_a^* \psi_b \inner{A}{B} + e^{-i\phi}\psi_a\psi_b^* \inner{B}{A} \; .
\end{equation}
A ``good'' measurement is normally taken to be one in which $a$ and $b$ can be distinguished with certainty, i.e., $\inner{A}{B}=0$; clearly, in this case, the interference terms vanish.  The first two terms are unchanged (since $A$ and $B$ are normalised), and the total probability becomes the incoherent sum of $|\psi_a|^2$ and $|\psi_b|^2$ as per rule (II).

But this formalism allows us to treat the more general situation, in which our measurement provides {\it some} information about which path was followed, but not perfect information.  In this case, the visibility is reduced, to a maximum value of $\left|\inner{A}{B}\right|$, which runs from $1$ in the case of no information (identical final environment states) to $0$ in the case of perfect information.  This is the origin of the ``duality relations'' 
studied by Greenberger and Yasin; Jaeger, Shimony, and Vaidman; Englert; and others, which can be written
\begin{equation}
\label{EG}
D^2 + V^2 \leq 1 \; ,
\end{equation}
where $D$ is the ``distinguishability,'' defined as $|P_A-P_B|$, the absolute value of the probability ``bias'' achieved between $a$ and $b$ by making the measurement; and $V$ is the visibility as described above.  To relate $D$ to $\inner{A}{B}$, it is sufficient to harken back to section \ref{USD}.  What is the minimum error probability for distinguishing states $A$ and $B$?  We saw that it was $\frac{1}{2}\left(1-\sqrt{1-\left|\inner{A}{B}\right|^2}\right)$.  It can be shown that this is equivalent to saying $D\equiv |P_A-P_B|=\sqrt{1-\left|\inner{A}{B}\right|^2}$, from which Eq.~\ref{EG} follows directly.

\subsection{The Quantum Eraser}
\label{QE}
During the development of quantum mechanics, it was taken for granted -- in practice, if not always in principle -- that a measurement had to involve amplification up to the macroscopic realm, and what Bohr described as an ``uncontrollable, irreversible disturbance.'' With the development of technologies such as cavity QED, however, it became clear that the sort of ``measurement apparatus'' we have just described could be a single photon, or other individual quantum particle.   Scully and various co-authors (notably Druhl, Hillery, Englert, and Walther) asked whether storing which-path information in a single quantum must necessarily disturb the system in the way Bohr imagined, and in particular, whether it might be possible to practically reverse it.  (While the Schr\"odinger equation is time-reversal symmetric, it is difficult to even conceive practically of reversing the motions of all the electrons, holes, and phonons excited in a slab of Silicon when a photon is detected by a classical avalanche photodiode: this is the same problem of practical irreversibility we come up against in classical statistical mechanics.  But if a measurement is effected by allowing a single atom to pass through a high-Q cavity for half a Rabi period, thus exchanging energy with a single mode of the radiation field, it is easy to imagine ``undoing'' this interaction in precisely the same manner.)  

\begin{figure}[tb]
\centering
\includegraphics[height=2in]{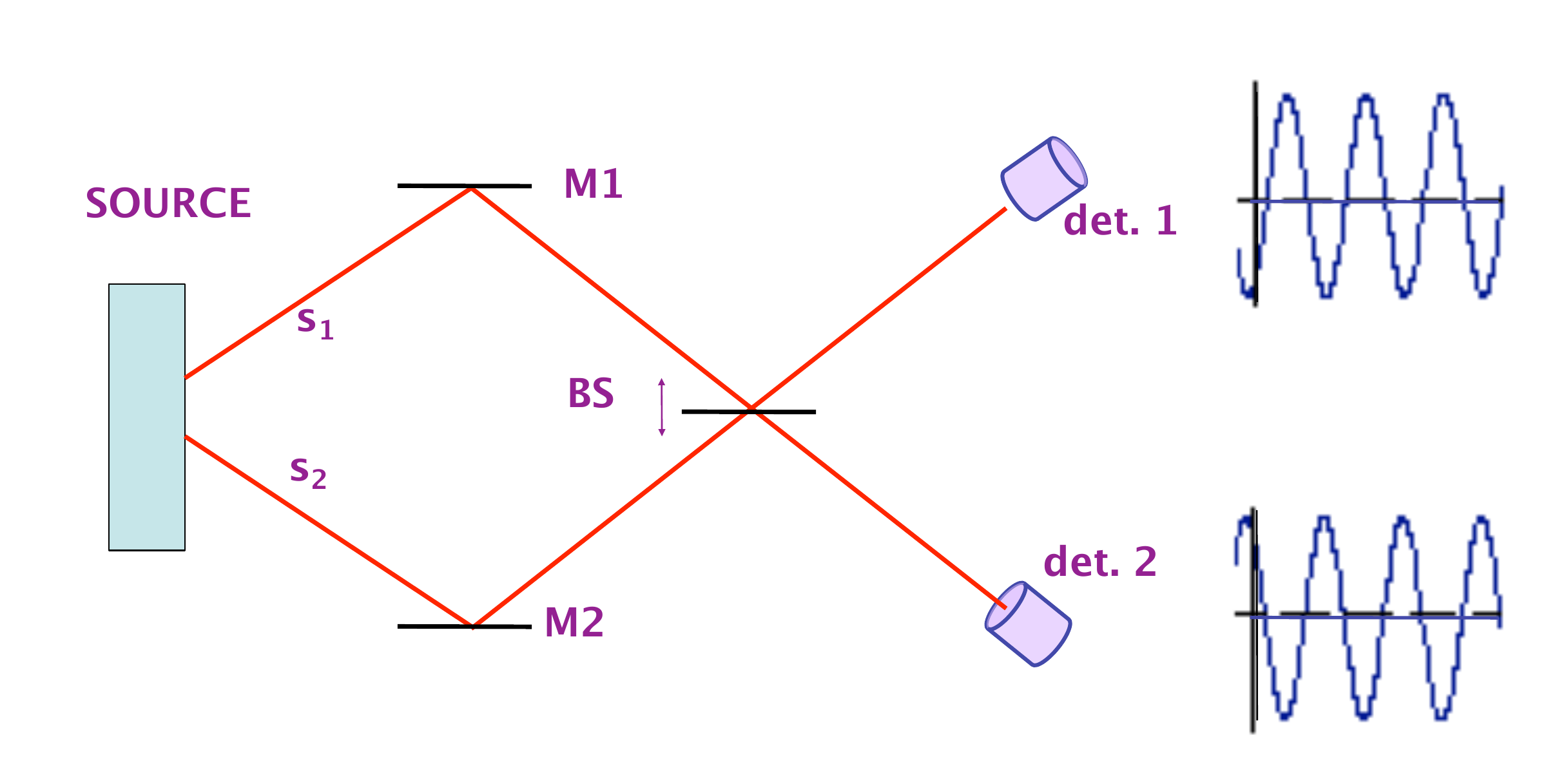}
\caption{A cartoon version of a single-photon interferometer: the source emits a ``signal'' photon in a superposition of paths $s_1$ and $s_2$, and the probability of each detector firing varies sinuosoidally (or cosinusoidally) with the phase difference between the two paths.}
\label{QE1}
\end{figure}

\begin{figure}[tb]
\centering
\includegraphics[height=2in]{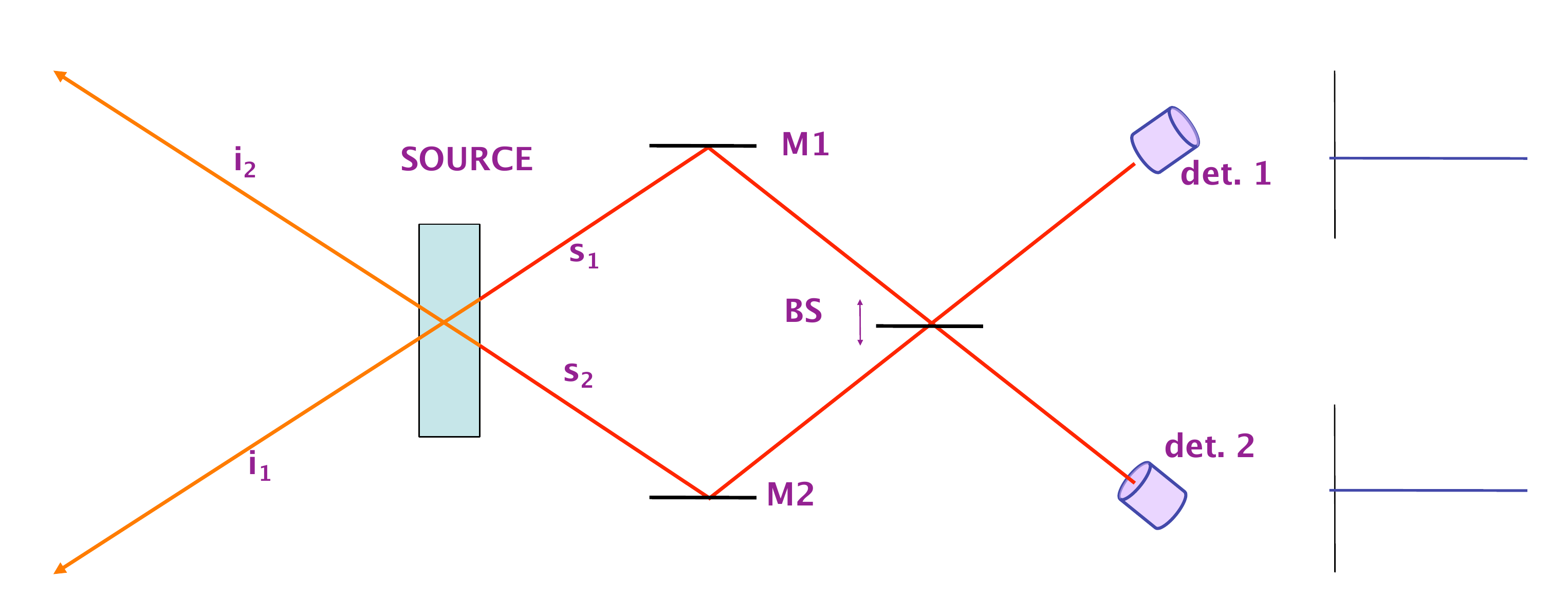}
\caption{Which-path information can be stored in an ``information photon'' entangled with our signal, such that the emitted state is $c_1\ket{s_1}\ket{i_1} + c_2\ket{s_2}\ket{i_2}$, as of course is the case in spontaneous parametric down-conversion.}
\label{QE2}
\end{figure}

As a toy example (whose connection to some of the classic experiments on entangled photons will easily become clear), imagine a source (Fig.~\ref{QE1}) which emits a single ``signal'' photon at a time, in a coherent superposition of paths $s_1$ and $s_2$ (this could simply be a beam splitter illuminated by a weak laser beam).  Interference is of course observed at the two detectors as the path lengths are changed, e.g. by displacing the beam splitter.  Now suppose that, unbeknownst to us, the NSA had convinced the manufacturer of the source to build in a ``back door,'' surreptitiously sending them an ``information photon'' $i_1$ whenever $s_1$ was emitted, or $i_2$ whenever $s_2$ was emitted.  This information photon of course plays the role of measuring apparatus, and if -- as in Fig.~\ref{QE2} -- $\inner{i_1}{i_2}=0$, our interference will be destroyed (and we will learn of the presence of the eavesdropper).  

\begin{figure}[tb]
\centering
\includegraphics[height=2in]{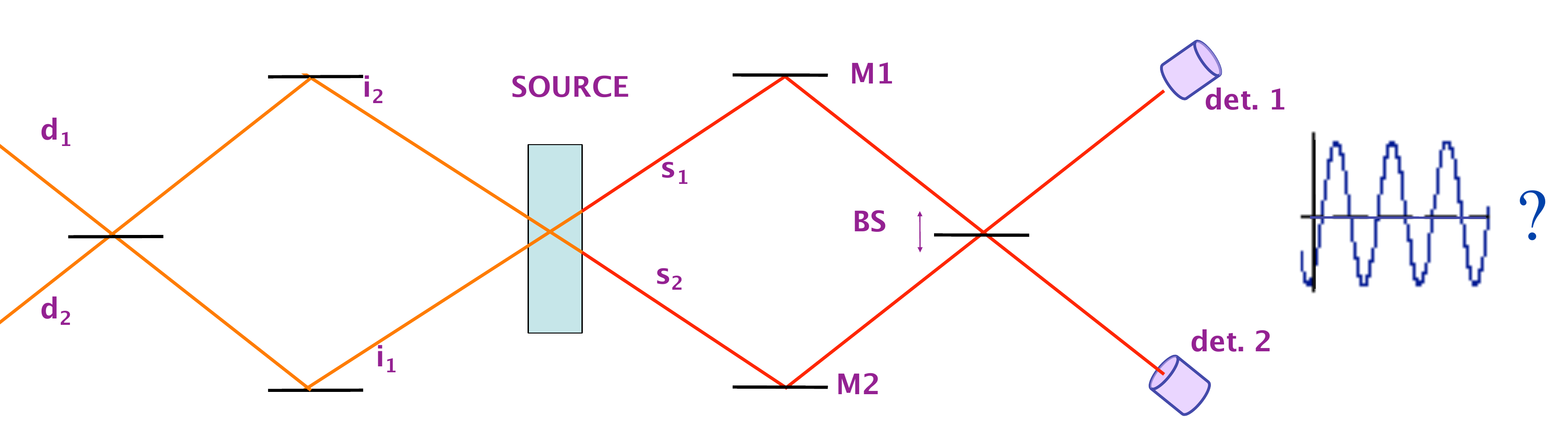}
\caption{Can the information carried by the ``i'' photons be easily erased?.}
\label{QE3}
\end{figure}

Scully {\it et al.}'s question was: if one could ``erase'' this information, making it impossible for any one to determine which path our signal had taken, would interference be restored?  To address this, let us think about how such erasure could take place.  A simple proposal is shown in  Fig.~\ref{QE3}.  If the two `i' paths are combined at a 50/50 beam splitter, then detecting a photon exiting either output port of the beam splitter would provide no information at all about whether it originated along $i_1$ or $i_2$.  Would this do the job?  If so, consider the consequences.  The ``erasure'' beam splitter could be added at any point in time; hypothetically, long after I had detected my signal photon, after the information photon had been propagating along one of its two paths for years.  An alien residing in the vicinity of $\alpha-$Centauri could decide at the last minute whether to keep the information (destroying the interference) or to erase it (restoring the interference).  And somehow, my dusty lab book from 4 years in the past would now have a record of interference or its absence, depending on what this alien had just done.  Obviously, this is impossible, and it is one way to see that {\it no} unitary evolution on the measuring apparatus can ever change the status of the signal interferometer.  Mathematically, this is because unitary evolution preserves inner products ($\bra{A}U^\dagger \;U\ket{B} = \inner{A}{B}$), so adding the beam splitter (which of course only leads to a unitary evolution of the photon modes) cannot change the quantum distinguishability of the paths, nor therefore the visibility.

Specifically, if the beam-splitter mixes $i_1$ and $i_2$ to create two new modes $d_1$ and $d_2$, it will in fact map each of them onto a different superposition:
\begin{eqnarray}
\ket{i_1} & \rightarrow & t\ket{d_1}+r\ket{d_2} \nonumber \\
\ket{i_2} & \rightarrow&  t\ket{d_2}+r\ket{d_1} \; ,
\end{eqnarray}
where unitarity demands that $r^*t + t^*r = 0$, and hence that the two final states are orthogonal.  This means that they are distinguishable in principle.  And in fact, it is easy to see that they are distinguishable in practice -- it suffices to add yet {\it another} beam splitter to recombine $d_1$ and $d_2$, such that they form two paths in a balanced Mach-Zehnder interferometer.  Then $i_1$ will be guaranteed to exit one port of the resulting interferometer and $i_2$ the other; placing detectors at those two ports would have precisely the same effect as placing them just behind the source.

The missing element in the functional quantum eraser was hinted at twice above.  On the one hand, no {\it unitary} evolution can change the overlap, but non-unitary evolution could (recall the solution to the unambiguous state discrimination problem).  On the other, if the ``erasure beam-splitter'' is inserted, then ``detecting a photon exiting [along path $d_1$ or $d_2$]  would provide no information at all.''  If a photon is actually {\it detected} in one of $d_1$ or $d_2$, then in some sense it is too late to add the extra beam splitter and distinguish between $\left(\ket{d_1}\pm\ket{d_2}\right)/\sqrt{2}$.  Well, strictly speaking, even this is not true.  If by ``detected,'' we simply mean the field modes have interacted with some measuring apparatus, then remember that in the state of the whole universe (including the measuring apparatus), no coherence has been lost, and it is {\it mathematically} conceivable that Schr\"odinger's cat could be interfered back together again.  The only way to guarantee that this does not happen is to {\it project} the system onto $d_1$ or $d_2$.  This is the non-unitary operation which allows the overlap to be modified.  Physically, it corresponds to the act of ``post-selection,'' i.e., studying only events where a given detector fired on the left, and discarding the others.

Let us analyze this in the framework of Eq.~\ref{MA}.
\begin{equation}
\frac{1}{\sqrt{2}}\left(\ket{s_1}+e^{i\phi}\ket{s_2}\right) \rightarrow \frac{1}{\sqrt{2}}\left(\ket{s_1}\ket{i1} + e^{i\phi}\ket{s_2}\ket{i_2}\right)
\end{equation}
exhibited no interference terms, because $\inner{i1}{i2}=0$.  But now let us {\it project} this state onto a particular equal superposition of $i_1$ and $i_2$, e.g., $\bra{d1} \equiv \left(\bra{i1}-i\bra{i2}\right)/\sqrt{2}$ for a symmetric 50/50 beam-splitter.  
\begin{equation}
\frac{\bra{i1}-i\bra{i2}}{\sqrt{2}} \left( \frac{\ket{s_1}\ket{i1} + e^{i\phi}\ket{s_2}\ket{i_2}}{\sqrt{2}} \right) = \frac{\ket{s_1}-ie^{i\phi}\ket{s_2}}{2} \; ,
\end{equation}
a resulting state which will of course exhibit interference (albeit with a $-\pi/2$ phase shift).  The non-normalized nature of this state reflects the fact that the post-selection only succeeds $50\%$ of the time.

Note that if one instead postselected $d_2$, by projecting onto $\bra{d2} \equiv \left(\bra{i1}+i\bra{i2}\right)/\sqrt{2}$, the resulting system state would be $\propto \frac{\ket{s_1}+ie^{i\phi}\ket{s_2}}{2}$, with a $+i$ in place of the $-i$; interference would still occur, but $180^\circ$ out of phase with the pattern observed for a $d_1$ postselection.  This comes back to the point established earlier: even after the ``erasure beam-splitter,'' the full ensemble displays no interference.  It is necessary to project out a {\it subensemble}, and only after this non-unitary projection can it be possible for interference to be restored.

The symmetry of the situation should be apparent.  What is really observed are {\it correlations} between the detectors for the information (or ``idler'') photon and the detectors for the signal photon.  For a given value of $\phi$, detecting a photon at $d_1$ will increase the probability of detecting a photon at the upper signal detector; for a different value of $\phi$, it will do the reverse.  But one can equally well argue that in the former case, detection at the upper signal detector increased the probability of detecting a photon at $d_1$.  The interference occurs not for subsystems but for the entire composite system, determining the probabilities of the four possible outcomes $s_1d_1$, $s_1d_2$, $s_2d_1$, and $s_2d_2$.  We are free to treat one subsystem or the other as a ``measuring apparatus,'' but there is no distinct physical role for this in the theory.

\section{``Interaction-free'' measurement and the trouble with retrodiction}
\label{IFM}

Perhaps the most remarkable example of a case of ``indirect'' measurement is what was colloquially known as the ``Elitzur-Vaidman bomb'' problem until political correctness (perhaps related to the fact that some of the experiments on this topic were carried out at Los Alamos when it was trying to shed its association with bombs) led to the seemingly more neutral name `` `interaction-free' measurement'' (IFM), which in turn proved so controversial that I shy away from typing it without extra scare quotes.  Of course, we have already recognized that every measurement begins with an interaction, but the phrase ``interaction-free'' here is meant to have a very specific sense.  To understand it, is is helpful (although by no means essential) to return to the original setup of the problem.  Imagine that you have purchased a supply of bombs which come equipped with triggers so sensitive that the passage of a single particle of {\it anything} would be guaranteed to set off an explosion -- a single molecule of air, a single electron, a single photon would be enough.  Now suppose that you are told that some of your stockpiled bombs have defective triggers, and will never blow up at all -- but that the {\it only physical difference} between the working bombs and the defective bombs is that the former will blow up, and the latter will not.  If this is the only physical difference, then it should seem intuitively clear that the only way to establish which kind of bomb one has is to try to set it off, and observe whether or not it explodes; this would make it impossible, of course, to establish that a given bomb was functional without destroying that bomb.  (Analogous peacetime applications could be considered, such as detecting the presence of a piece of unexposed film without exposing the film; or taking an X-Ray of a subject without running the risk that the subject will be harmed by the absorption of the X-Rays; but for the present, we will content ourselves with the theoretical ramifications of the problem.)  The goal Elitzur and Vaidman set themselves was to determine whether or not a bomb was operational -- at least some of the time -- without destroying it.  And since interaction with even a single photon was hypothesized to trigger any working bomb, this task can be considered in that limited sense ``interaction-free.''

\begin{figure}[tb]
\centering
\includegraphics[width=4in]{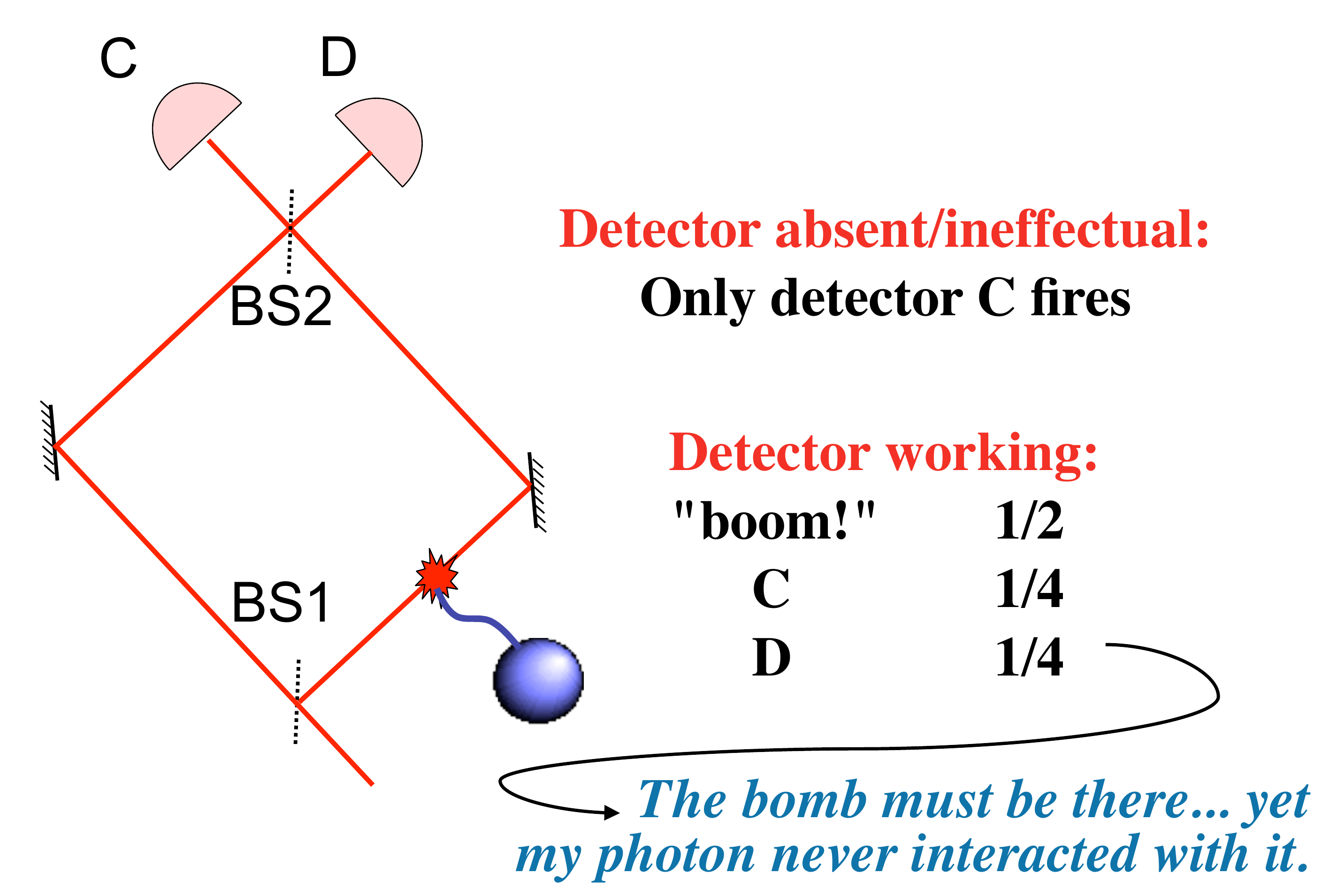}
\caption{Interaction-free measurement proceeds by placing the object to be studied in one path of a balanced Mach-Zehnder interferometer, which in the absence of any disturbance exhibits constructive interference at detector $C$ and destructive at detector $D$, making the latter a ``dark port.''}
\label{IFM-fig}
\end{figure}

While the classical intuition is faultless classically, and the task is indeed impossible, in quantum physics the situation is different.  The setup can be seen in Fig.~\ref{IFM-fig}.  A balanced (equal-path-length) Mach-Zehnder interferometer is built, and illuminated with one photon at a time.  Due to constructive interference, every photon reaches detector $C$, making detector $D$ a ``dark port.''  Now the trigger of a bomb is placed in one path of the interferometer.  For simplicity, let us suppose the trigger detector is of the QND (quantum non-demolition) sort, that is, that the photon passes through unscathed, whether or not the trigger works.  If this is a working trigger, then the bomb serves as a ``measuring apparatus'' for the photon; if the photon follows the right-hand path, the bomb explodes, and if it follows the left-hand path, it does not.  Presumably, these two final states have vanishing overlap, and if my lab is still in one piece, I can safely conclude that the photon took the left-hand path.  But of course, if I have which-path information, no interference is possible: therefore, one half of the photons which avoid the bomb ($25\%$ overall) reach detector $D$.  On the other hand, if the trigger had been defective, no which-path information would have been obtained, and interference would persist: no photons would have reached $D$ at all.  Armed with this knowledge, I conclude that whenever a photon reaches $D$, this indicates the presence of a working bomb (one with the capability of establishing which-path information), {\it even} when (because I did not blow up) I am certain that the photon did not follow the path containing the bomb.  

This is one of those straightforward consequences of quantum interference which are so striking that quantum optics researchers, when first told about it, seemed to split fairly evenly into a camp which called it impossible and a camp which called it trivial.  It is of course related to the fact that quantum mechanics is not a theory of events, or even of probabilities, but of probability {\it amplitudes}.  In some sense, Feynman's injunction that we must add the amplitudes for every path which could possibly have been followed means that it is not a theory of ``what happened,'' but of ``all the things that could have happened.'' Thus the question of whether or not the photon {\it could} have taken the other path without exploding a bomb is a salient one.  To those who simplistically feel that quantum mechanics proves the old chestnut that ``a tree falling in a forest when there is no one there makes no sound,'' I would counter that this effect demonstrates that sometimes, we can tell whether or not the tree {\it would} have made a sound, even without felling it.

Of course, only $25\%$ of the time do we confirm that the bomb is working, without destroying it.  $25\%$ of the time, we get no information; and $50\%$ of the time, we need to replace our destroyed laboratory.  Well, this is not quite true.  First of all, what can we conclude if detector $C$ fires?  We know that
\begin{equation}
P(C|{\rm working}) = \frac{1}{2} \; {\rm and} \; P(C|{\rm defective}) = 1 \; ;
\end{equation}
we remember from Eq.~\ref{Bayes} that an observation of $C$ will therefore double the ratio of $P({\rm defective})$ to $P({\rm working})$; if the two possibilities were equally likely, then a single photon at $C$ makes the probability of a defective bomb $2/3$, two photons make it $4/5$, three photons make it $8/9$, and so on.  In the case of a defective bomb, we can send as many photons through the interferometer as we like, to reach a desired level of certainty that it is in fact defective.  For a working bomb, on the other hand, we can also send another photon through every time one lands at $C$, but each time, the bomb will be twice as likely to blow up as to divert the photon to $D$.  Thus if we continue sending photons through the system until one of the (more or less) conclusive results is reached, $2/3$ of the working bombs will blow up and $1/3$ will be ``detected.''  By modifying the reflectivities of the beam splitters appropriately, Elitzur and Vaidman showed that this success probability could be raised arbitrarily close to $1/2$ -- and Kwiat {\it et al.} later showed that by using a ``Zeno-effect''-based strategy (in practice, a multi-pass rather than a single-pass interferometer) one could in principle identify arbitrarily close to $100\%$ of the bombs without detonating them!

\subsection{Hardy's ``Retrodiction Paradox''}
\label{hardy}
Clearly, there is still an interaction {\it Hamiltonian} at work in the IFM, even if in specific cases, our final state may escape the nefarious influences of this interaction.  But there is a deeper concern here.  It is fundamental to quantum mechanics that measurement (at least of a quantity which was not already in a definite eigenstate) changes the state of the system.  Elitzur and Vaidman and many others tried to study the case of a {\it quantum} bomb, and how this indirect measurement would affect it, if the measurement could proceed without blowing it up.  The analysis due to Hardy has \begin{figure}[tb]
\centering
\includegraphics[width=4in]{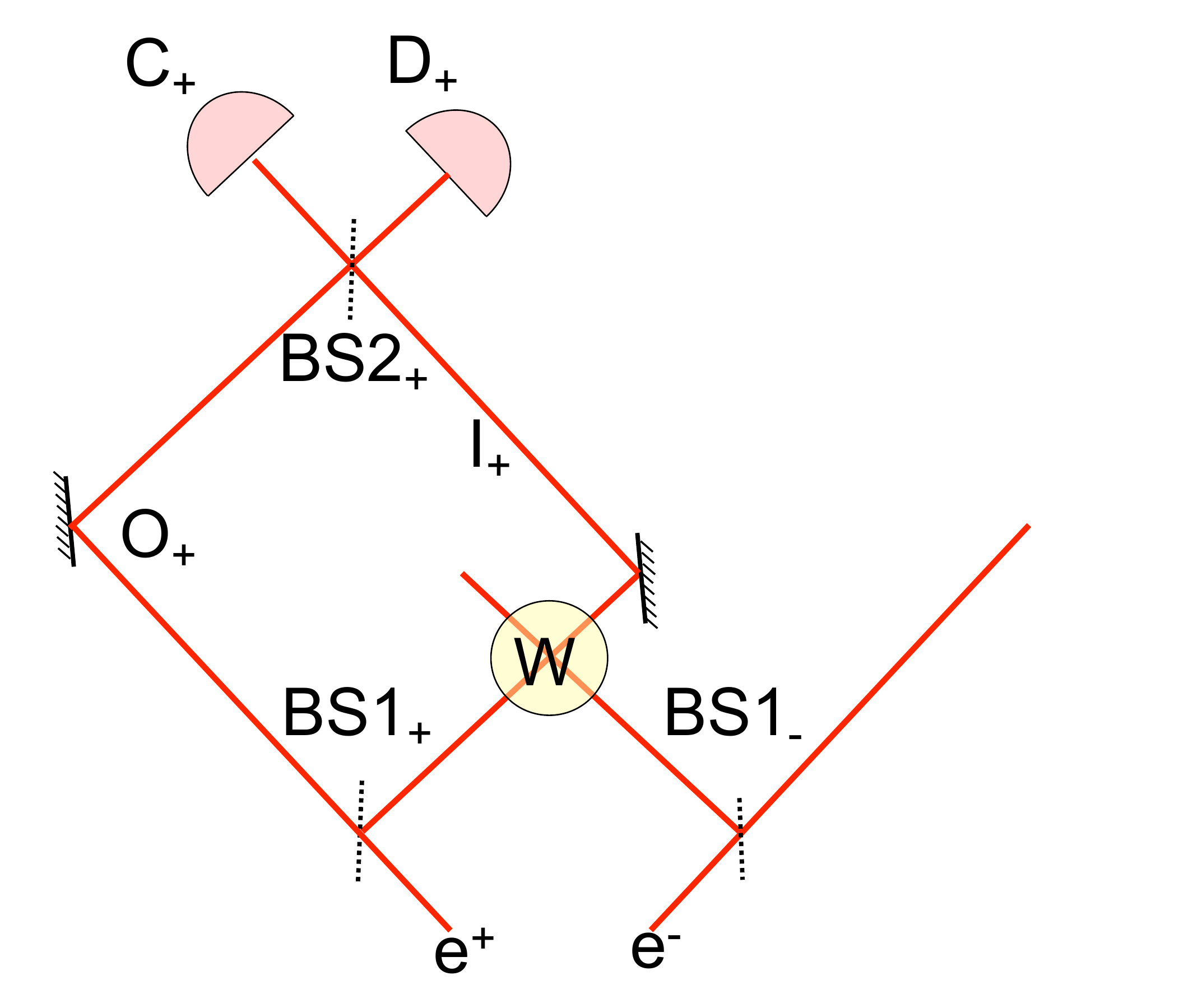}
\caption{Hardy's version of IFM for a quantum object: an electron which is in a superposition of being in ``interaction region'' $W$ and being outside.  If an electron and positron meeting at $W$ are certain to annihilate, then a positron reaching detector $D_+$ might lead us to conclude that the electron was in fact ``in.''}
\label{Hardy1}
\end{figure}
had the greatest long-standing impact.  Instead of considering a bomb which could be functional or defective, Hardy considered a single electron in a superposition of being in the interferometer and being elsewhere.  In his Gedankenexperiment, the photon interferometer was replaced with a positron interferometer (see Fig.~\ref{Hardy1}), and it was assumed to be arranged such that if the electron met the positron in region $W$, they would be certain to annihilate.  In this way, the electron in $W$ functions as a working bomb, while the electron in an outside path plays the role of the defective bomb.  According to the logic of IFM, the positron can only reach detector $D_+$ if the electron is in fact in $W$.  But this should ``collapse'' (or at least decohere) the state of the electron -- something which could be studied by adding an additional electron beam splitter, to ``close the interferometer'' for the electron, as in figure \ref{Hardy2}

\begin{figure}[tb]
\centering
\includegraphics[width=4in]{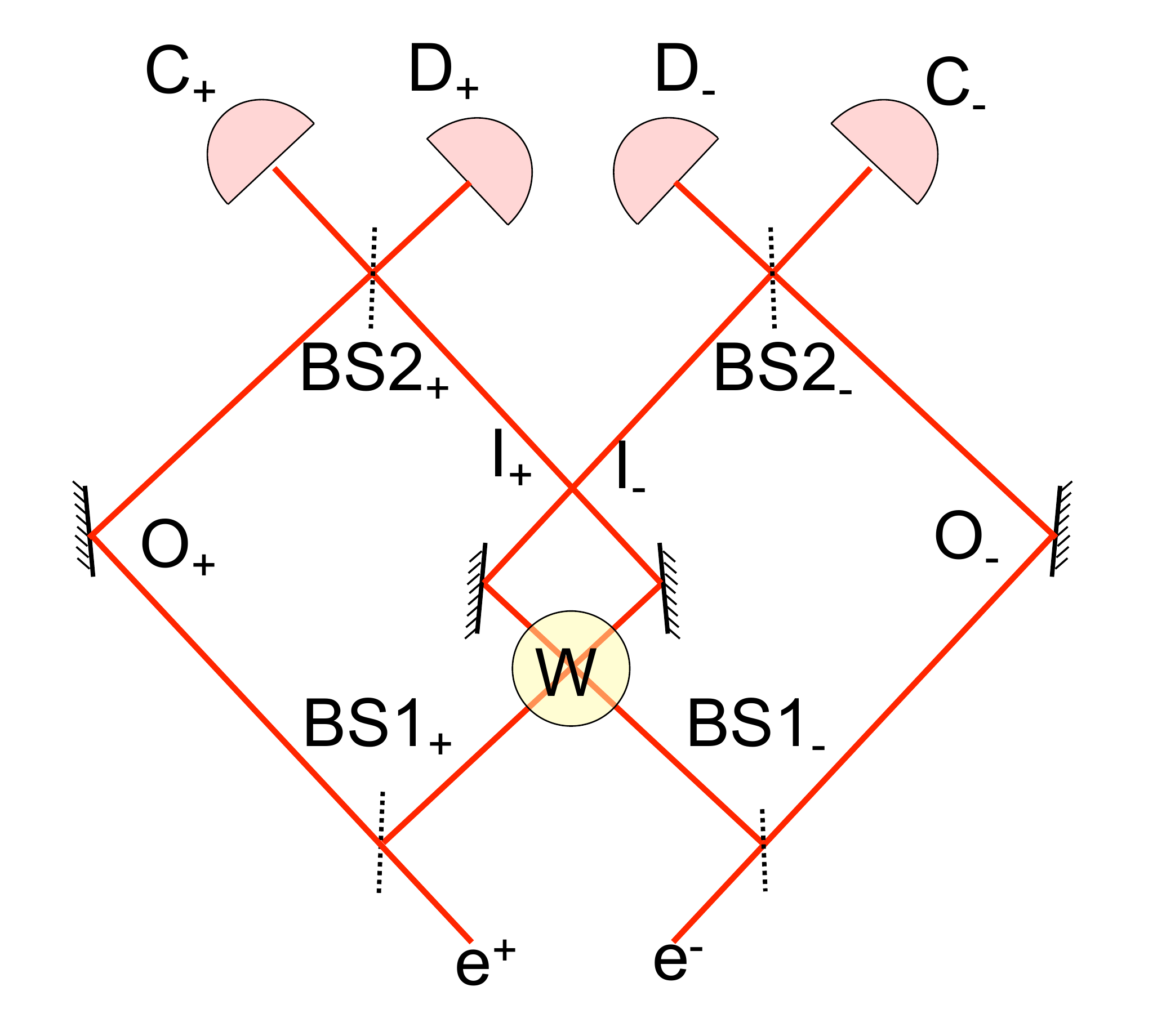}
\caption{If detection at $D_+$ tells us the electron was in $W$ while detection at $D_-$ tells us the positron was in $W$, yet the electron and positron annihilate whenever they meet in $W$, can the two $D$ detectors ever fire simultaneously?}
\label{Hardy2}
\end{figure}

A glance at this figure should immediately convince you that the IFM discussion suggests that $D_+$ can only fire if the electron is in $W$, but since the {\it electron} is in its own balanced interferometer, $D_-$ can only fire if the {\it positron} is in $W$.  At first, this seems borne out by direct calculation.  Let us start with an initial state
\begin{equation}
\ket{\psi_i} = \frac{\ket{O_+}+\ket{I_+}}{\sqrt{2}} \otimes \frac{\ket{O_-}+\ket{I_-}}{\sqrt{2}}  \; .
\end{equation}
After the interaction in $W$, this state has become
\begin{equation}
\label{boom}
\ket{\psi_f} = \frac{\ket{O_+}\ket{O_-} + \ket{I_+}\ket{O_-} + \ket{O_+}\ket{I_-}}{2} + \frac{\ket{\rm boom}}{2} \; .
\end{equation}
Since $\ket{C+}$ is meant to be the port with fully constructive interference (prior to any electron-positron interaction), we define $\ket{C+}\equiv\left(\ket{O_+}+\ket{I_+}\right)/\sqrt{2}$, and $\ket{D_+}$ must then be the orthogonal state $\left(\ket{O_+}-\ket{I_+}\right)/\sqrt{2}$.  If we detect the positron at detector $D_+$, the state of the electron can be determined by projecting onto this:
\begin{equation}
\inner{D_+}{\psi_f} = \frac{\ket{O_-}-\ket{O_-}+\ket{I_-}}{2\sqrt{2}} = \frac{\ket{I_-}}{2\sqrt{2}} \; .
\end{equation}
What do we conclude?  First, that the probability of $D_+$ firing is $|1/2\sqrt{2}|^2$ or $1/8$ -- this is the 50\% chance that the electron was in $W$, times the $25\%$ chance of a ``successful interaction-free measurement'' on the first try.  Second, we conclude that whenever $D_+$ fires, the electron is indeed in the state $I_-$.  So far, so good.  But now let us ask what the probability of $D-$ firing on the same occasion is.  Intuitively, an electron in $I_-$ has a $50\%$ chance of reaching detector $D-$, so we see immediately that the overall probability of $D_+$ {\it and} $D_-$ firing on the same occasion is $1/16$, or
\begin{equation}
\left|\bra{D_-}\inner{D_+}{\psi_f}\right|^2 = \left|-\frac{1}{4}\right|^2 \; .
\end{equation}

But wait.  $D+$ could only fire if the electron was in $W$; $D-$ could only fire if the positron was in $W$; and if the electron and positron were both in $W$, they are sure to annihilate, and therefore no detector at all should fire.  Yet a few-line calculation shows us that one time in 16, this paradoxical event occurs.  This leads many to the conclusion that there is something fundamentally wrong with ``interaction-free'' measurements: in essence, that although there may be cases in which you reach the correct conclusions, by avoiding the physical effects of the interaction, you have side-stepped one of the necessary conditions for a true measurement, and cannot be sure that your conclusions will always be valid.  Hardy attributes this to the feature of quantum mechanics known as {\it contextuality}.  Due to the theorem of Kochen and Specker, it is known to be impossible to assign ``non-contextual'' values to observables.  That is, if you wished to construct a ``hidden-variable theory'' consistent with quantum mechanics, you could not assume that even on individual events, the value you would get if you measured some observable $A$ would be independent of whether you measured $A$ along with some other operators $B$ and $C$ or whether you measured it along with $B^\prime$ and $C^\prime$ -- even if all the operators in either set $(A,B,C)$ or $(A,B^\prime,C^\prime)$ are known to commute and hence be compatible.  The result of a measurement of $A$ depends on the {\it context}.  Thus the question of whether the electron was in $W$ or not could have a definite answer if you measured it directly -- but a different answer if you looked for the electron at $D_-$ instead.  

To many, this paradox reinforced the traditional view of quantum measurement.  You cannot measure something without disturbing it; and if we did not directly measure which path the electron took through its interferometer, we cannot assign a particular history to it based on observations of the positron; we cannot ``retrodict'' its behaviour in this indirect fashion.  In the following section, I will introduce another modern measurement paradigm which takes precisely the opposite view, and then examine how they may be reconciled.

\section{Strong and weak measurements: from von Neumann to Aharonov}

\subsection{von Neumann's measurement interaction}
\label{vN}

Throughout these lectures, we have followed the view that a measurement proceeds by allowing a system to interact with some ``probe,'' after which we can examine the state of the probe.  This is precisely how von Neumann himself modelled the physical process of measurement.  He imagined a ``pointer,'' like the needle on a dial, whose position after a measurement interaction could depend on some observable of the system.  In order to treat this interaction quantum mechanically, his pointer of course had to be a quantum system.  Let us consider it to be described by some wave function $\psi(x_p)$, where $x_p$ represents the pointer position.  This position has some associated uncertainty $\Delta x_p$, but for the measurement to be useful, we would assume that $\Delta x_p$ is not too big; an ideal ``classical'' measurement would arise in the limit $\Delta x_p \rightarrow 0$.

\begin{figure}[tb]
\centering
\includegraphics[width=4in]{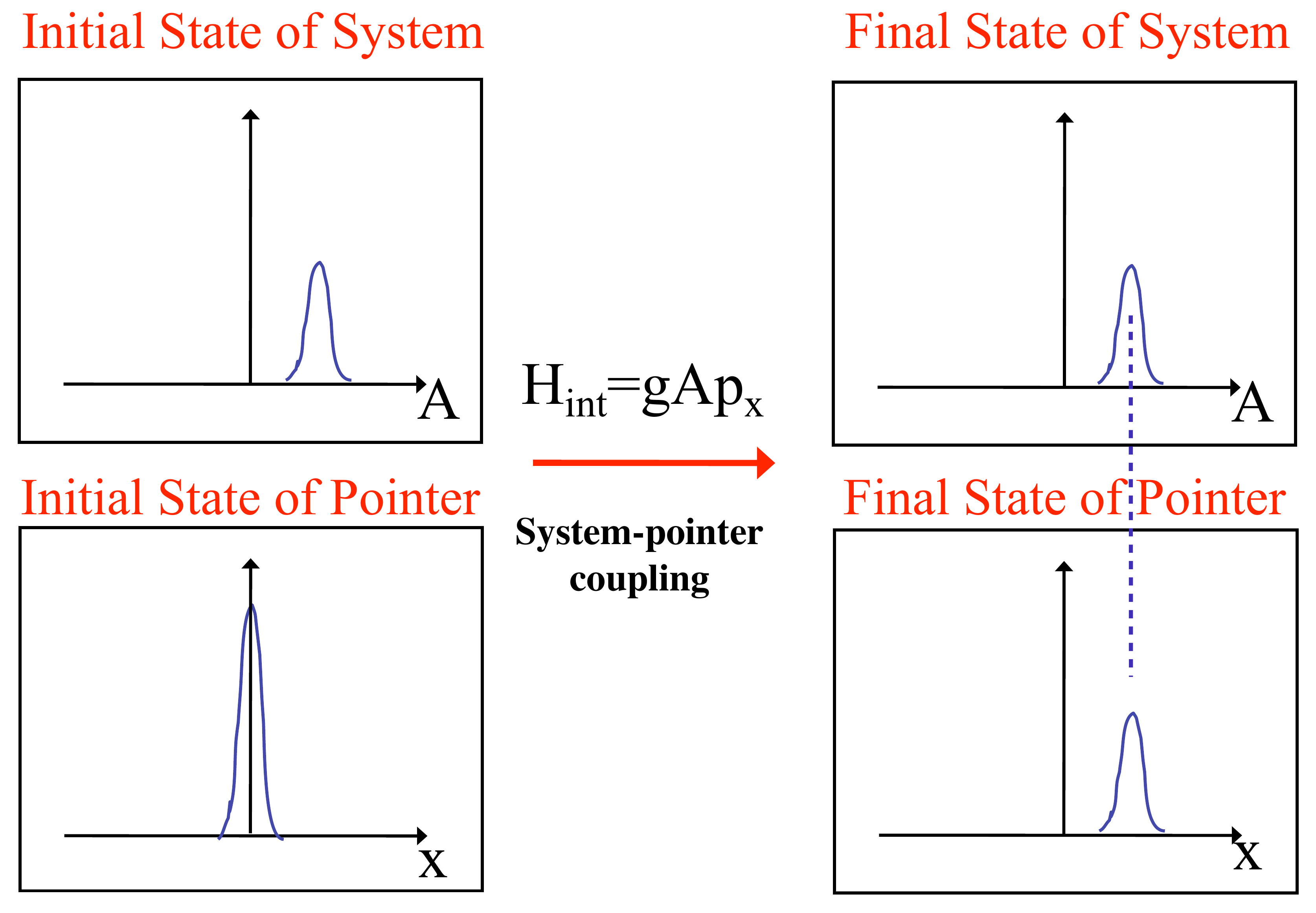}
\caption{von Neumann's hypothetical measurement interaction couples the system observable of interest, $A_s$, to the momentum $p_x$ of the pointer.  Since the momentum is the generator of translations, the net effect is to translate the pointer wave function from $\expect{x_p}=0$ to $\expect{x_p}\propto \expect{A_s}$. }
\label{vN1}
\end{figure}

In order to measure a system observable $A_s$, we would like to design an interaction such that the pointer position -- initially assumed to be $\expect{x_p}=0$ -- becomes proportional to the value of $A_s$.  Since the pointer momentum operator $P_p$ is the generator of translations, this can be accomplished by using an interaction Hamiltonian ${\cal H}_{\rm int} = g(t)A_sP_p$, where $g(t)$ is some time-dependent interaction strength.  This leads to
\begin{eqnarray}
\frac{d}{dt}\expect{x_p} &=& \frac{1}{i\hbar}\expect{[x_p,{\cal H}]} \nonumber \\
& = & \frac{g(t)}{i\hbar}\expect{A_s}[x_p,P_p] = g(t)\expect{A_s} \; .
\label{evol}
\end{eqnarray}
If we assume that the interaction is brief enough that no time-evolution of $A_s$ itself need be considered, we find immediately for the pointer shift
\begin{equation}
\label{shift}
\Delta\expect{x_p} = \int dt g(t) \expect{A_s} \equiv G \expect {A_s} \; ,
\end{equation}
where we have defined $G$ as the integrated interaction strength.

The bottom line is that if system and pointer are coupled through a product of their respective operators, the system operator is the observable {\it being measured}; while the pointer operator is the {\it conjugate} (``pointer momentum'') to the observable which will serve as a readout (``pointer position'').  Some find the situation more intuitive by flipping the roles.  One could measure $A_s$ by having the system exert a force on the pointer proportional to the value of $A_s$.  Since a constant force is a potential $U(x_p) = -Fx_p$, this would be accomplished by a Hamiltonian ${\cal H}_{\rm int} = -(dF/da)Ax_p$.  Here the proportionality constant $dF/da$ is the interaction strength, and $x_p$ now plays the role of ``pointer momentum.''  How can we see this?  If this Hamiltonian acts on the system and pointer for some time, the applied force will impart a momentum shift to the pointer, which is proportional to the value of $A_s$.  Thus it is the {\it momentum} of the physical pointer which can be used to read out the system observable, and which we would typically refer to as the ``pointer position.''

\begin{figure}[tb]
\centering
\includegraphics[width=4in]{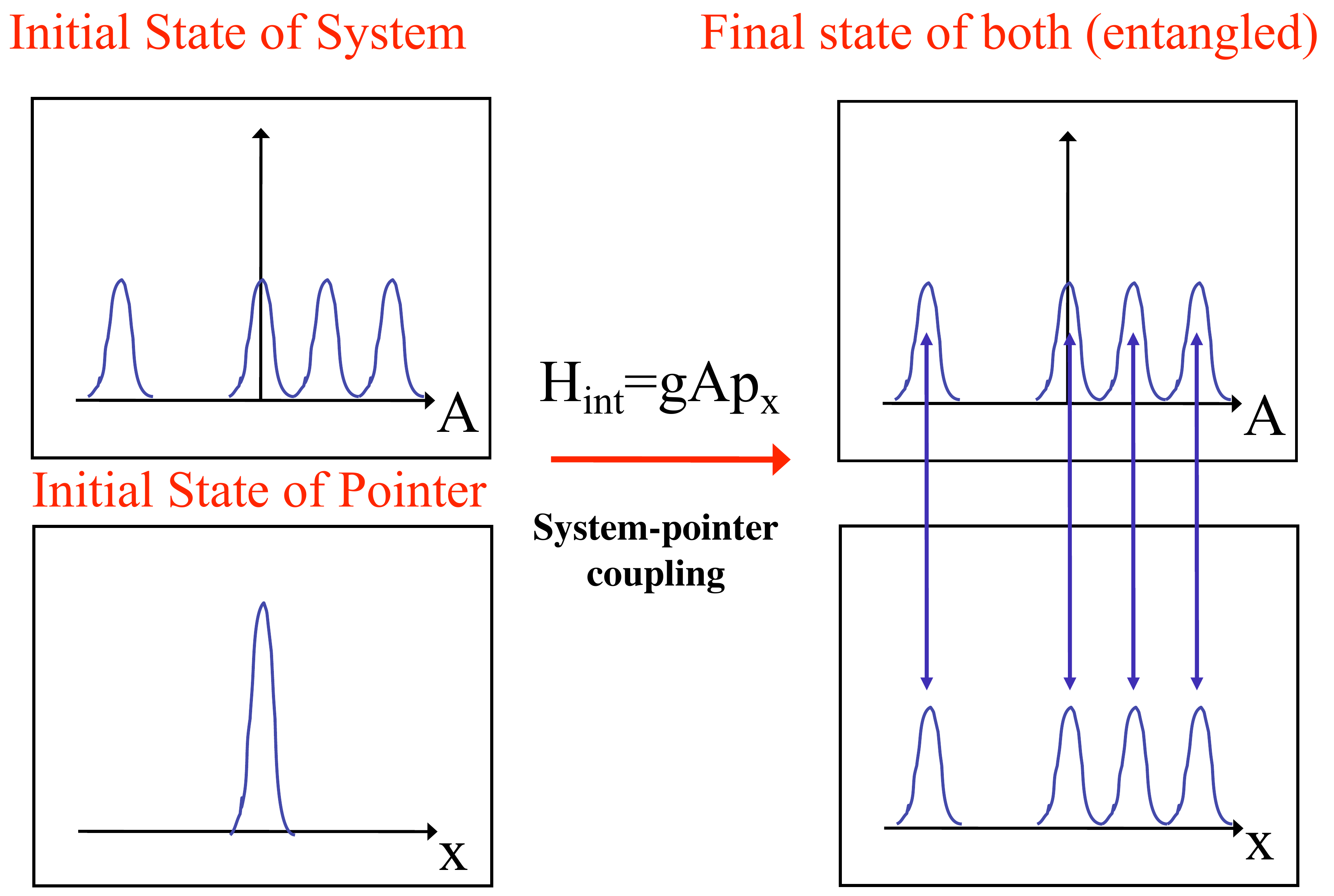}
\caption{If the characteristic pointer shifts are large compared to the initial width of the pointer wave function, then in general the final state of system and pointer will be entangled, each value of $A_s$ correlated with an appropriately shifted pointer state. }
\label{vN2}
\end{figure}

Note that if $A_s$ is in an eigenstate, the operator in ${\cal H}_{\rm int}$ can be replaced by the corresponding eigenvalue $a$, and the Hamiltonian becomes a pure displacement operator.  The pointer wave function is shifted, without distortion, by an amount $Ga$.  But if the system is initially in a superposition of states with different eigenvalues $a_i$, then by the superposition principle, system+pointer evolve into a superposition in which each $A_s$-eigenstate is correlated with an appropriately displaced pointer:
\begin{equation}
\label{entang}
\left\{c_1\ket{a_1} + c_2\ket{a_2}+c_3\ket{a_3}+\ldots\right\}\psi(x_p) \\
\; \; \rightarrow c_1\ket{a_1}\psi(x_p-Ga_1) + c_2\ket{a_2}\psi(x_p-Ga_2) +c_3\ket{a_3}\psi(x_p-Ga_3)+\ldots \; .
\end{equation}
The typical assumption, following von Neumann, is that the scale of the position shifts ($Ga_i$) is much larger than the initial uncertainty of the pointer position $\Delta x_p$; otherwise, it would not be possible to determine the value of $A_s$ from the pointer.  If this is the case, then the pointer wave functions in the above expression are nearly orthogonal, making it a highly entangled state of system and pointer, represented schematically in Fig.~\ref{vN2}.  As in the discussion of Eq.~\ref{MA2}, such a measurement completely destroys the coherence between difference eigenstates of $A_s$.  The measurement has disturbed the system; in particular, by changing the phase relationship between $A$-eigenstates, it has disturbed the observable conjugate to $A_s$ (as changing the phase relationship between different positions disturbs the momentum, and changing the phase relationship between different momenta disturbs the position).

There is a complementary way to understand the origin of this disturbance.  In the limit of a ``good'' measurement, we assumed $\Delta x_p$ had to be small.  This means, of course, that $\Delta P_p$ must be large.  But from the point of view of the system, the interaction Hamiltonian which acts on it through $A_s$ is proportional to $P_p$; if this latter quantity is uncertain, then the system feels an effect of uncertain magnitude.  Since $A_s$ is the displacement operator for its own conjugate variable, the disturbance manifests itself on this conjugate.  ($A_s$ itself commutes with ${\cal H}_{\rm int}$ by construction, and can therefore not be disturbed by this interaction.)  For instance, if $A_s$ were the position of some particle, the Hamiltonian would correspond to an actual force, proportional to $P_p$, acting on the particle.  The more uncertain $P_p$, the more uncertain the force, and the greater the random disturbance to the particle's final momentum (the variable conjugate to $A_s$).  The conclusion is already well-known to you.  The more accurately you measure $A_s$ (e.g., particle position), the more disturbance you create for the conjugate variable (e.g., particle momentum).  Note that in von Neumann's formalism, this result falls out directly of the coherent Schr\"odinger interaction between system and pointer, and no assumption of any ``state reduction'' is required.  Mathematically, the ``randomness'' arises when we trace over the pointer with which the system has become entangled, leaving the latter in a mixed state with reduced coherence.

\subsection{Weak measurement}
\label{weak}

Aharonov and his coworkers realized in the late 1980s that this formalism opened the door to a new possibility.  If, contrary to von Neumann, we take the opposite limit, of large $\Delta x_p$, then it is possible to have a state with small $\Delta P_p$, and hence little disturbance arising in the interaction Hamiltonian.  Seen from the perspective of Eq.~\ref{entang}, the large position uncertainty means that the overlap between $\psi(x-Ga_i)$ and $\psi(x-Ga_j)$ may be nearly unity, so that very little information is gained from an individual measurement and there is very little concomitant 
loss of coherence (recall section \ref{comp}).  
(The state very nearly factors into a product of a system state and a pointer state, so that tracing over the latter leaves the former unchanged.)  Note that the {\it average} shift of the pointer is still given by $G\expect{A_s}$, as we saw in Eq.~\ref{shift}, which did not require any assumptions about the width of the pointer wave packet.  

Part of the motivation for thinking about such ``weak'' measurements, where the disturbance could be reduced at the expense of the amount of information extractable, was to enable one to discuss measurements on post-selected systems, such as those we encountered in the discussion of the quantum eraser, or indeed Hardy's paradox.  Could we imagine measuring which path the electron had been in, but only on those occasions where it eventually reached a particular detector?  Of course, the standard view of quantum mechanics would have us say no; we must choose whether to place a detector inside the interferometer or after the final beam splitter.  But by now we realize that the choice between ``a detector'' and none is too limiting; a measurement can be any interaction between the system and some second object, and this interaction may be stronger or weaker.  The time-reversibility of the Schr\"odinger equation led Aharonov to argue that one should be able to get as much information about present observables from final conditions (post-selection) as from initial conditions (state preparation).  In the usual paradigm, the symmetry is apparently broken by the ``uncontrollable, irreversible'' nature of traditional measurement.  I cannot really talk about the electron which was following a trajectory from a given input to a given output, if at some intermediate time I measure where it is, violently disturbing its momentum.  But, argued Aharonov {\it et al.}, in the limit of an arbitrarily weak measurement, the disturbance to my system could be arbitrarily small.  Of course, the deeper one moves into this limit, the smaller the effect on the pointer.  However, one could repeat identical measurements on a large ``subensemble'' of identically prepared {\it and} identically post-selected systems, and the average pointer shift would still provide information.  (As observed above, in the absence of post-selection, the pointer shift remains proportional to $\expect{A}$, no matter how small the coupling $G$ or how large the pointer uncertainty $\Delta x_p$.)

\begin{figure}[tb]
\centering
\includegraphics[width=4in]{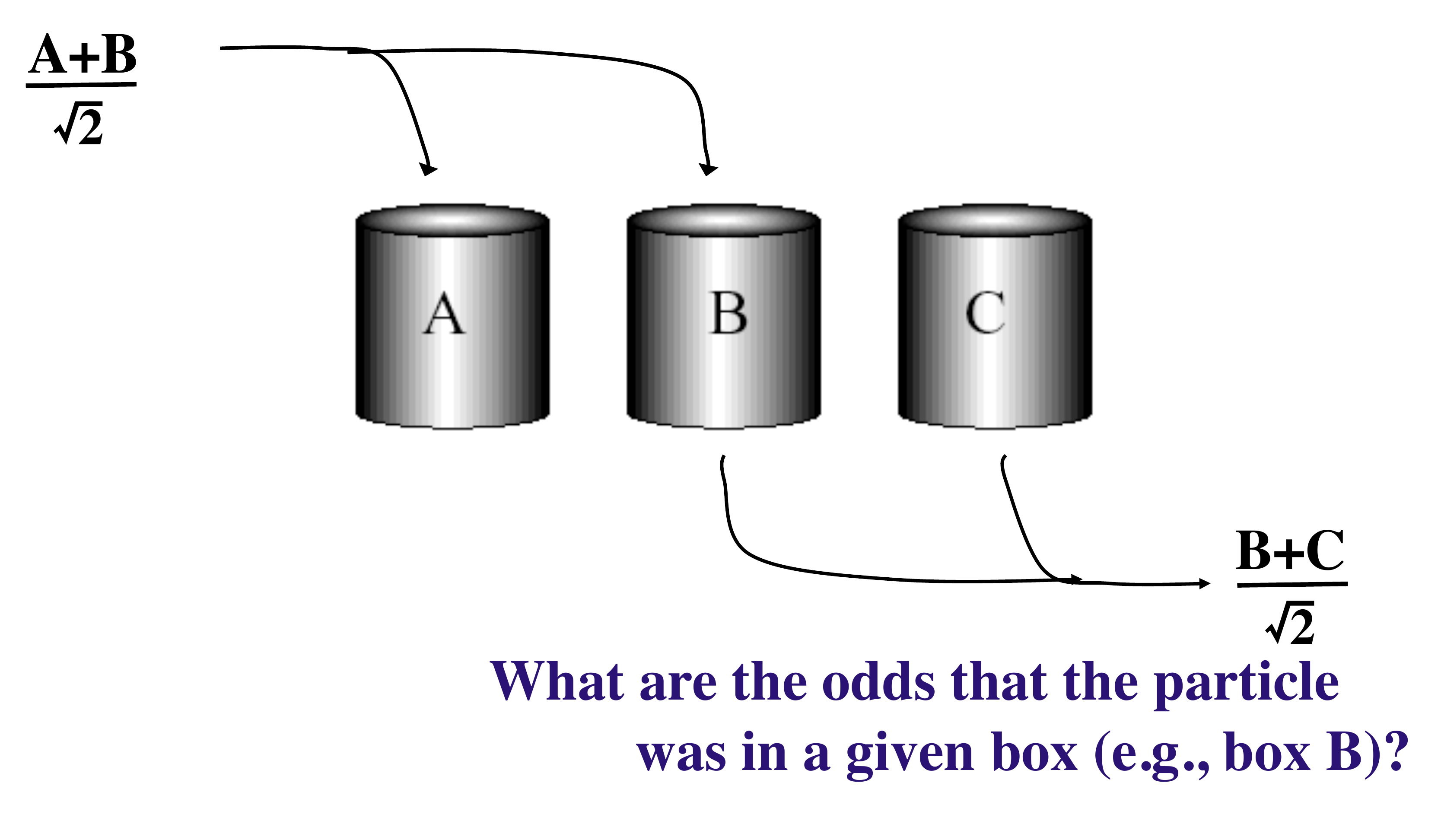}
\caption{The quantum 3-box problem.   If I promise you that I will prepare a \$20 bill in the state $\left(\ket{A}+\ket{B}\right)/\sqrt{2}$ and that I will find it in the state $\left(\ket{B}+\ket{C}\right)/\sqrt{2}$, how much can you conclude about where it is likely to have been between the preparation and the post-selection?  Would it be possible to verify this?}
\label{3Box}
\end{figure}

Figure \ref{3Box} shows a particular example.  I show you three boxes, labelled $A$, $B$, and $C$, and offer to play a game.  I will hide a \$20 bill in the boxes, and if you can find the bill, you get to keep it.  But to make the game more interesting, instead of placing the bill in a particular box, I place it in a coherent superposition $\left(\ket{A}+\ket{B}\right)/\sqrt{2}$.  To make the game more fair, on the other hand, I give you an additional piece of information: I promise that I will find it in {\it another} superposition, $\left(\ket{B}+\ket{C}\right)/\sqrt{2}$.  (Of course, I cannot guarantee that this will occur on every try; it is merely a post-selection.  I promise you that if I {\it fail} to find the bill in that superposition, we will simply set up the game again and you will have a second chance, and so on until the post-selection succeeds.)  Which box do you bet on?

Most physicists when asked this question refuse to answer; we have been trained to believe that the bill does not reside in any particular box, but really is in a superposition of $A$ and $B$ until I choose to measure whether or not it is in $\left(\ket{B}+\ket{C}\right)/\sqrt{2}$, at which instant it makes a ``quantum jump'' into that state, or an orthogonal one.  Most non-physicists immediately give the right answer, which is $B$.  There is no way the bill could have been in $C$, given the promised preparation (we neglect the possibility of tunneling between boxes); yet there is no way the bill could have been in $A$, in light of the state I found it in.  According to this rather classical reasoning, there is a $100\%$ probability that the bill was in $B$.  But are we allowed to do this quantum mechanically or not?  Is ``$\ket{A}+\ket{B}$'' tantamount to saying ``A or B,'' or not?  The very fact that physicists hesitate when presented with this problem seems to me to point to a gap in our real intuition for the quantum world, and this is at least one important role for discussions of weak measurements.

For indeed, if -- before the post-selection was attempted -- you made a projective measurement of whether or not the bill was in one of the boxes, you would greatly disturb the state of the bill.  There would still be correlations with the success of the postselection, but you might not wish to think of the bill you observed as being in the process of following its natural trajectory from preparation to postselection.  Nevertheless, the correlations are easy to reason through: if you found the bill in box $A$, the post-selection could never succeed.  And naturally, you could never find the bill in box $C$ in the first place.  Therefore, if you looked in $A$, $B$, and $C$ on various occasions, the only place you could find the particle {\it and} have the post-selection succeed would be box $B$.  Not finding the particle in $B$ would collapse it into $A$ and guarantee that the post-selection would fail.  Thus even for projective measurements, in this case the classical intuition holds.  More generally, the results of strong measurements conditioned on later post-selection are given by the ABL (Aharonov-Bergmann-Lebowitz) rule, which is essentially the straightforward application of Bayesian reasoning to these inferences, along the lines that we have applied several times now in these lectures.  In what follows, though, we will consider the results of {\it weak} measurements on such systems, asking what the average shift of a weakly coupled pointer would be, on those occasions when the post-selection succeeded.

This is of interest because there are more general situations in which one can really not tolerate the disturbance of these intermediate strong measurements, unlike the case of the 3-box problem.  For instance, consider a particle which is at $t=0$ in a well-localized, real-valued wave packet centered at $x=0$.  Its momentum is of course very uncertain, and has zero expectation value due to the real value of the wave function.  But suppose that at $t=\tau$ you make a projective measurement, finding the particle in a well-localized wave packet centered at $x=x_0$ (also with a very uncertain momentum).  What would you conclude about the average value of the momentum at times $0<t<\tau$?  The strict textbook approach rejects the question; or rather, suggests that $\expect{P}$ was 0, and remained 0 until the instant when the final measurement disturbed the system (after which $\expect{P}$ was in fact 0 again anyway).  Common sense, on the other hand, suggests that a particle which travelled a distance $x_0$ in a time $\tau$ should have had a momentum somewhere around $mx_0/\tau$.  But of course, if a strong measurement of momentum were actually made between times $0$ and $\tau$, any momentum within the initial uncertainty could be found; and since any momentum eigenstate is delocalized, the post-selection would be equally likely regardless of which was obtained. It turns out in such cases that if a {\it weak} measurement is performed, the average result conforms with intuition; this is a result of a number of mathematical properties of weak measurements that make some researchers wish to ascribe some deep significance to the values they reveal.  Such an attribution remains controversial, however, so we will focus on the incontrovertible experimental predictions of the formalism, which rely only on standard quantum theory.

Suppose a system begins in some initial state $\ket{i}$, and a pointer in some state $\ket{\psi_p}$.  We once more assume an interaction Hamiltonian ${\cal H}= g(t)A_sP_p$, and an integrated interaction strength $G\equiv\int g(t) dt$, but we will now make the additional assumption that the measurement is weak, i.e., that the characteristic pointer shifts |GA| are small compared to the smallest length scale set by the pointer momentum distribution, $\hbar/|P|$.  Schr\"odinger evolution yields
\begin{eqnarray}
U\ket{\psi_p}\ket{i} & = & e^{-\frac{i}{\hbar}\int dt {\cal H}(t)}\ket{\psi_p}\ket{i} \nonumber \\
& \approx & \left(1-\frac{i}{\hbar}GA_sP_p\right)\ket{\psi_p}\ket{i} \; .
\label{GAP}
\end{eqnarray}
Now suppose we successfully post-select the particle to be in final state $\ket{f}$.  The state of the pointer can be determined by projecting Eq.~\ref{GAP} onto $\bra{f}$:
\begin{eqnarray}
\label{disp}
\ket{\psi_p^\prime} & = & \bra{f}U\ket{\psi_p}\ket{i} \approx \inner{f}{i}\ket{\psi_p} - \frac{iG}{\hbar}\bra{f}A_s\ket{i}P_p\ket{\psi_p} \nonumber \\
& = & \inner{f}{i} \left[1-\frac{iG}{\hbar}\frac{\bra{f}A_s\ket{i}}{\inner{f}{i}} P_p\right]\ket{\psi_p} \nonumber \\
& \approx &  \inner{f}{i} e^{-iGa_wP_p/\hbar}\ket{\psi_p} \; ,
\end{eqnarray}
where we have defined
\begin{equation}
\label{WVA}
a_w \equiv \frac{\bra{f}A_s\ket{i}}{\inner{f}{i}} \; ,
\end{equation}
and made the weakness assumption that $|GA_sP_p| \ll \hbar$.

Recalling that $P_p$ is the generator of translations, we see that the action of the exponential in Eq.~\ref{disp} is to translate $\ket{\psi_p}$ by the amount $Ga_w$.  Recalling that (in the absence of post-selection) the pointer shift was $G\expect{A_s}$, we see that this shift corresponds to the result we would expect if the value of $A$ were given by $a_w$, which is termed the ``weak value,'' and given by Eq.~\ref{WVA}.  

\subsection{Some implications of the weak-value formula}

The question of whether or not to ascribe ``reality'' to this weak value between the preparation and the postselection is a difficult philosophical one.  But it has a rigorously defined operational meaning, in that it tells us how strong the effect of the system would be on any other object coupled to $A_s$, in the limit where this coupling was not so large as to significantly disturb the evolution of the system between preparation and postselection.

Some interesting properties are immediately evident from Eq.~\ref{WVA}.  If either the initial or the final state is an eigenstate of $A_s$, then the weak value is guaranteed to be the corresponding eigenvalue: we get information as reliably from final conditions as from initial conditions.  Furthermore, the formula is linear; the weak value of $A+B$, generally denoted $\expect{A+B}_{\rm wk}$, is equal to $\expect{A}_{\rm wk}+\expect{B}_{\rm wk}$, regardless of whether or not $A$ and $B$ commute.  (Roughly speaking, this is why from information about $\expect{X(0)}_{\rm wk}$ and $\expect{X(\tau)=X(0)+P\tau/m}_{\rm wk}$  -- the localization of a particle at two instants in time -- we can draw conclusions about $\expect{P/m}_{\rm wk}$, the velocity of the particle.)  In a sense, this points to one of the most dramatic features of weak measurements, which is their non-contextuality.  While the Kochen-Specker theorem tells us that we cannot imagine assigning definite values to particular observables on a shot-by-shot basis without knowledge of the ``context'' of the measurement, the weak-value formula tells us that the weak value of a given observable in some post-selected ensemble is entirely independent of any other weak values that might be measured simultaneously.  As we shall see, this makes weak measurement a particularly interesting approach from which to reanalyze Hardy's paradox, and its suggestion that retrodiction (the stuff of IFM, but really the stuff of weak measurement as well) is inherently untrustworthy.

It is also simple to show that if post-selection is done onto a complete, orthonormal set of final states, and the weak value is averaged over all these post-selections, weighted by the success frequency, one recovers the usual expectation value.  This has to be the case, since if one sums up all the possible post-selections, one is back to measuring the pointer shift irrespective of the final state of the particle, as in Eq.~\ref{shift}.  On the other hand, Eq.~\ref{WVA} has some disquieting features as well.  For instance, as $\inner{f}{i} \rightarrow 0$, the weak value may diverge.  In fact, the weak value is free to take essentially arbitrary values, even ones entirely outside the eigenvalue spectrum of $A_s$.  Of course, given the large pointer uncertainty assumed at the outset, there is never an individual event on which one can say that one observed a value of $A_s$ beyond its spectrum; but the average shift could be extremely large.  Since this only occurs when the postselection probability $\left|\inner{f}{i}\right|^2$ is very small, there is no contradiction with the behaviour of the full ensemble, for which the pointer shifts by an amount $G\expect{A_s}$ which is of course constrained by the eigenvalues of $A_s$.  

Worse, $a_w$ is not even guaranteed to be real, even for Hermitian operators $A_s$.  But there is a straightforward interpretation of an imaginary part to the weak value.  The effect of 
\begin{equation}
e^{-iG(i \, {\rm Im} \, a_w) P_p/\hbar}\ket{\psi_p}
\end{equation}
is to multiply the momentum-space wave function $\tilde{\psi}(p) \equiv \inner{p}{\psi_p}$ by a weighting factor $\exp[(G\,{\rm Im}\,a_w)p/\hbar$.  Unlike the real part of $a_w$, which indicates a position shift for the pointer (the expected result of a measurement), this imaginary part occasions a {\it momentum} shift of the pointer.  For the simple case of an initial Gaussian pointer wave function, this is easy to calculate:
\begin{eqnarray}
e^{-x^2/4\sigma^2} & \Rightarrow & e^{-(x-Ga_w)^2/4\sigma^2} \nonumber \\
&  \propto & e^{-(x-G\, {\rm Re} \,a_w)^2/4\sigma^2}e^{(ixG\,{\rm Im}\,a_w)/2\sigma^2} \; ,
\end{eqnarray}
where the first exponential indicates a position shift for the pointer of $\Delta x = G\, {\rm Re} \,a_w$ and the second is a momentum boost of $\Delta p = G\,{\rm Im}\,a_w / 2\sigma^2$.  Note that unlike the position shift, the momentum shift depends on the width of the initial pointer wave packet; as the measurement is made infinitely weak and $\sigma \rightarrow \infty$, this effect vanishes.  It is a reflection of the {\it back-action} of the measurement on the system.  Recall that when $a_w$ is summed over all possible post-selections, it recovers the expectation value.  Since $\expect{A_s}$ is real, this means that the imaginary parts cancel out; only when a particular post-selection is made can this change the expectation value of the pointer momentum, and for the simple reason that the presence of the pointer momentum in the interaction Hamiltonian could disturb the particle, making the post-selection more or less likely to succeed for different values of momentum (within the initial distribution).  As $\sigma$ gets larger and larger, the pointer momentum distribution may be made narrower and narrower, until in the limit, the back-action goes away.

What are the implications for the 3-box problem?  Measuring whether or not the bill is in box $B$ is simply measuring the projection operator $\proj{B}$; the eigenvalues of this operator are $1$ (for ``yes, the bill is in $B$'') and $0$ (for ``no, the bill is elsewhere''), and its expectation value represents the probability that the bill is in $B$.  Given 
\begin{equation}
\ket{i} = \frac{\ket{A}+\ket{B}}{\sqrt{2}}
\end{equation}
and
\begin{equation}
\bra{f} = \frac{\bra{B}+\bra{C}}{\sqrt{2}} \; ,
\end{equation}
it can be seen at once that $\bra{f}{\rm Proj}(B)\ket{i} = 1/2$ and $\inner{f}{i} = 1/2$, implying that $\expect{{\rm Proj}(B)}_{\rm wk} = 1$: the weak-valued probability is $100\%$, meaning on an operational level that when the post-selection succeeds, the effect of an interaction on a pointer would be {\it exactly} as strong as if the bill had simply been prepared entirely in box $B$.

The three-box problem discussed by Aharonov {\it et al.} is more surprising even that this.  Recall that the only property essential to proving that we could not find the particle in $A$ or $C$ was the orthogonality of states $A$ and $C$.  But let us then make three new boxes, $A^\prime$, $B$, and $C^\prime$, and replace $A$ with $A^\prime+C^\prime$ and $C$ with the orthogonal state $A^\prime-C^\prime$.  Specifically, suppose the bill is prepared in
\begin{equation}
\ket{i} = \frac{\ket{A^\prime}+\ket{B}+\ket{C^\prime}}{\sqrt{3}}
\end{equation}
and post-selected in
\begin{equation}
\ket{f} = \frac{\ket{A^\prime}+\ket{B}-\ket{C^\prime}}{\sqrt{3}} \; .
\end{equation}
Now, the particle cannot be found in $\ket{A^\prime}-\ket{C^\prime}$, as that is orthogonal to $\ket{i}$; nor can it be found in $\ket{A^\prime}+\ket{C^\prime}$, as that is orthogonal to $\ket{f}$.  Once more, the only remaining state is $\ket{B}$, and we conclude that it must be found there with certainty.  But a moment's inspection will alert you to the fact that both $\ket{i}$ and $\ket{f}$ are symmetric under exchange of $A^\prime$ and $B$.  The same argument then applies to $A^\prime$.  The particle cannot be found in $\ket{B}+\ket{C^\prime}$, nor can it be found in $\ket{B}-\ket{C^\prime}$, so it must be found in $\ket{A^\prime}$.  Indeed, it is easy to calculate $\expect{P_{A^\prime}}_{\rm wk} = \expect{P_{B}}_{\rm wk} = 1$.  But can a particle be in two places at once?  More to the point, we know by completeness that ${\rm Proj}(A^\prime) + {\rm Proj}(B) + {\rm Proj}(C^\prime) $ equals the identity $I$, with expectation value $1$.  So by linearity of weak values, $\expect{P_{A^\prime}}_{\rm wk}+\expect{P_{B}}_{\rm wk}+\expect{P_{C^\prime}}_{\rm wk} = 1$.  This relation can only be satisfied because $\expect{P_{C^\prime}}_{\rm wk}=-1$.  While these negative weak values for positive-valued operators may seem especially disturbing, their operational meaning is precisely the same as before.  If a pointer is allowed to interact with box $C$ in such a way that if the particle was in box $C$, the pointer would get attracted to the box, this negative weak value means that upon successful post-selection of the system, the {\it average} pointer would exhibit a momentum shift away from the box.  This is a real physical consequence of the weak values, and although I will not delve into it here, all such effects can be understood in terms of quantum interference.

\begin{figure}[tb]
\centering
\includegraphics[width=4in]{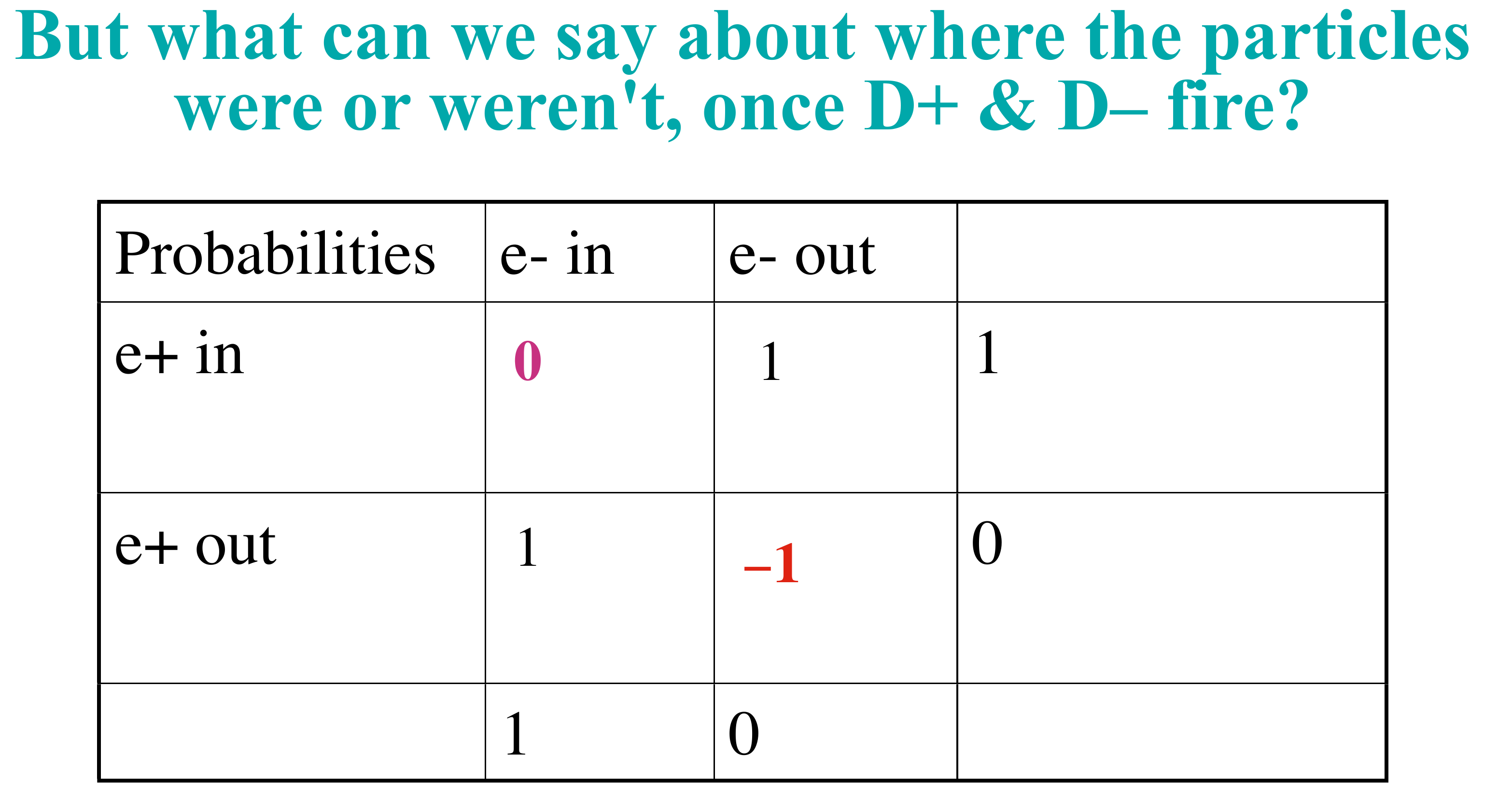}
\caption{The weak-value ``resolution'' to Hardy's paradox.  The weak-valued probability for both particles to be in $W$ simultaneously vanishes, even while the weak-valued probability for {\it each} of them to be in $W$ is $100\%$; the two statements can only be reconciled due to the possibility of negative weak-valued probabilities.}
\label{table}
\end{figure}

This brings us back to the question of Hardy's paradox, and whether or not one can use ``post-selections'' (detections of the positron at $C_+$ or $D_+$, for instance) to draw conclusions about the past whereabouts of the electron.  A few minutes' thought should suffice to convince you that indeed, when $D_+$ fires, the weak value of the projector onto $\ket{I_-}$ (the state in which the electron is in the interaction region $W$) is $1$.  In the {\it measurable} sense of the average effect on a weakly-coupled pointer, the retrodiction is completely valid.  And yet quantum mechanics tells us that $D_+$ and $D_-$ may fire together -- how is this possible?  When $D_+$ and $D_-$ both fire, does that mean that the weak-valued ``probability'' for the electron and the positron to be in $W$ simultaneously is $100\%$?  As a matter of fact, no.  Since the annihilation removed the term $\ket{I_+}\ket{I_-}$ from the superposition in Eq.~\ref{boom}, the weak value of the probability for both particles to be in $W$ is strictly $0$.  How, then, can the weak-valued probability for {\it each} particle to be in $W$ be $100\%$?  The resolution is of precisely the same spirit as the 3-box problem: there is $100\%$ weak-valued probability for the electron to be in $W$ and the positron to be outside, but there is {\it also} a $100\%$ weak-valued probability for the positron to be in $W$ and the electron to be outside; and since the complete set of projectors must add up to unity, there is a $-100\%$ weak-valued probability for {\it neither} particle to be in $W$.  The situation is summarized in Fig.~\ref{table}.

\begin{figure}[tb]
\centering
\includegraphics[width=4in]{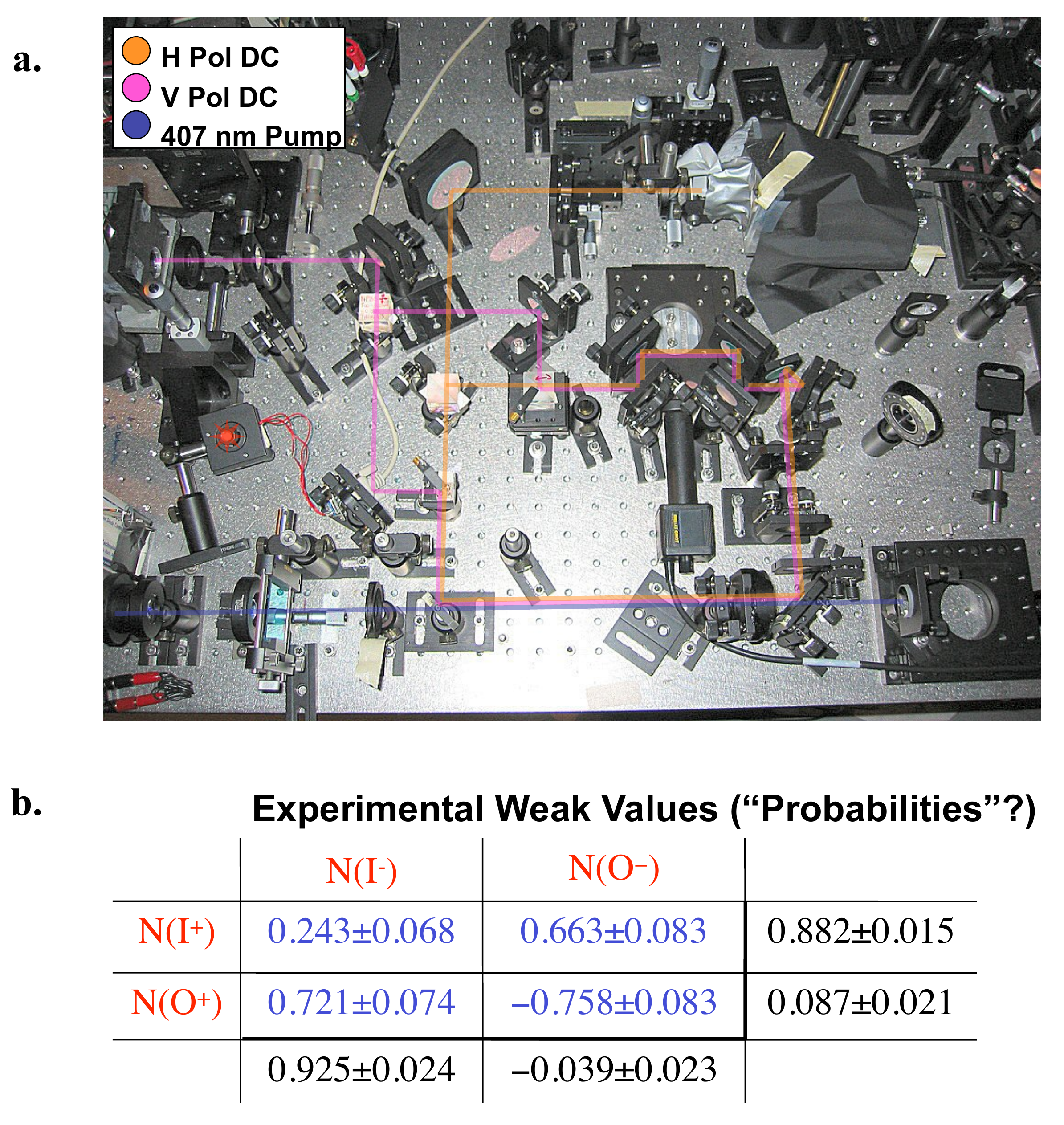}
\caption{The Toronto experiment on Hardy's Paradox.  (a) A photo of the two-photon interferometer which replaced the electron and positron interferometers of the original proposal, quantum-enhanced upconversion in a $\chi^{(2)}$ crystal playing the role of the $e^-e^+$ annihilation.  (b) Experimentally extracted weak values, to be compared with the table of figure \ref{table}..}
\label{Hardy-expt}
\end{figure}

These results were recently confirmed in two experiments (see Fig.~\ref{Hardy-expt}).  In conclusion, if one is willing to accept the statement that the value of an observable can only be determined in practice by measuring the size of its effect on some other physical system; and that if the interactions disturb the system, it is reasonable to use the weakest possible interaction strengths to minimize the disturbance, while measuring the {\it ratio} of the effect to the interaction strength; then these weak values provide a natural way to talk about properties of systems at times intermediate to preparation and post-selection.  Given the importance of post-selection in many aspects of quantum information, this should make them of particularly wide applicability.  And, despite the seemingly paradoxical nature of Hardy's thought experiment and its implications about contextuality, once one follows this set of definitions, one finds that there are no paradoxes, but rather a fully self-consistent and operational framework within which one can visualize the time-evolution of a quantum system.

\subsection{A Bayesian rederivation of the weak-value formula}
Given the way we introduced measurement, in terms of Bayesian inference, it may be interesting to see an alternate construction of the weak-value formula, based purely on conditional probabilities.  Starting from a reasonable proposal for what quantum-mechanical expression to write down to replace classical conditional probabilities, Eq.~\ref{WVA} can be derived in a few lines without any reference to the form of the von Neumann interaction, or Schr\"odinger evolution.  Along with the other tantalizing mathematical properties of weak values, this close connection to probability theory is one more hint that perhaps these expressions do have some fundamental significance in quantum mechanics.  Recall that a probability is given by the expectation value of a corresponding projection operator, with its eigenvalues of 0 and 1.  Let us make the assumption that the {\it joint} probability of two propositions $j$ and $f$ is given by $\expect{{\rm Proj}(f){\rm Proj}(j)}$, where since projectors onto different bases do not in general commute, we will choose the time-ordered product, with the earlier measurements on the right and the later ones on the left.

The expectation value of $A$ can be written
\begin{equation}
\label{expect1}
\expect{A} = \sum_j a_j P(j) \; ,
\end{equation}
where $a_j$ are the eigenvalues of $A$ and $P(j)$ are their respective probabilities.  If by a ``weak value,'' what we really mean is simply the ``conditional expectation value,'' i.e., the average value of $A$ {\it given} that a desired final state $\ket{f}$ is observed, then it is natural to write by analogy
\begin{equation}
\label{awk}
\expect{A}_{\rm wk} = \sum_j a_j P(j|f) \; .
\end{equation}
We already saw that
\begin{equation}
P(j|f) = \frac{P(j\& f)}{P(f)} \; .
\end{equation}
Given our joint-probability hypothesis, we can evaluate this as
\begin{equation}
P(j|f) = \frac{\expect{{\rm Proj}(f){\rm Prof}(j)}}{\expect{{\rm Proj}(f)}} 
\end{equation}
and plug this into Eq.~\ref{awk} to calculate 
\begin{eqnarray}
\expect{A}_{\rm wk} & = & \sum_ja_j  \frac{\expect{{\rm Proj}(f){\rm Prof}(j)}}{\expect{{\rm Proj}(f)}}  \nonumber \\
& = & \frac{\expect{{\rm Proj}(f) \sum_ja_j {\rm Prof}(j)}}{\expect{{\rm Proj}(f)}} \nonumber \\
& = & \frac{\expect{{\rm Proj}(f)A}}{{\rm Proj}(f)} = \frac{\inner{i}{f}\bra{f}A\ket{i}}{\inner{i}{f}\inner{f}{i}} \nonumber \\
&  =&  \frac{\bra{f}A\ket{i}}{\inner{f}{i}}
\end{eqnarray}

%
%

\section{Acknowledgments}

I would like to acknowledge all members of my group, present and past, for the collaborations which have informed
these notes, along with the many people who have influenced the development of my thinking about the topics treated in these lectures, notably Yakir Aharonov, Lev Vaidman, Sandu Popescu, J\'anos Bergou, Robin Blume-Kohout, and Howard Wiseman.

\newpage

 \thebibliography{0}
 
 {\bf FOR FURTHER READING}\\
 \vspace{0.2in}\\

{\bf More Detailed References Related to Quantum Measurement:} \\

Kurt Jacobs, {\it Quantum Measurement Theory and its applications}, Cambridge University Press, Cambridge, due out August 2014 (several sections available as free pdf download at
http://www.quantum.umb.edu/Jacobs/books.html ) \\

M. A. Nielsen and I. L. Chuang.Ê{\it Quantum Computation and Quantum Information},ÊCambridge University Press, Cambridge, 2000  \\

C. W. Helstrom.Ê{\it Quantum Detection and Estimation Theory},Êvolume 123 ofÊMathematics in Science and Engineering,ÊAcademic Press,ÊNew York, 1976. \\

Howard M. Wiseman,ÊGerard J. Milburn, {\it Quantum Measurement and Control}, Cambridge University Press, 2009.\\

John von Neumann, {\it Mathematical Foundations of Quantum Mechanics}, Princeton University Press (1955)\\

V.B. Braginsky, F.Ya. Khalili, and K.S. Thorne, {\it Quantum Measurement}, Cambridge University Press (1992)\\

\\

W. Zurek, ``Decoherence and the transition from quantum to classical," Physics Today 44, 36 (1991)\\

\\

A.M. Steinberg, lecture notes on ``Experimental Quantum Measurement,'' available online at http://www.physics.utoronto.ca/$\sim$aephraim/2206/\#notes \\

\vspace{0.2in}\\

{\bf State discrimination:} \\

I. D. Ivanovic, Phys. Lett. A 23 257 (1987).\\
A. Chefles and S. M. Barnett, J. Mod. Opt. 45, 1295 (1998)\\
S. M. Barnett and E. Riis, J. Mod. Opt. 44, 1061 (1997)\\
B. Huttner et al., Phys. Rev. A 54, 3783 (1996)\\
R. B. M. Clarke et al., Phys Rev A 63, 040305 (2001)\\
R. B. M. Clarke et al., Phys Rev A 64, 012303 (2001)\\
T. Rudolph, R. W. Spekkens, and P. S. Turner, Phys. Rev.
A 68, 0101301 (2003)\\
M. Takeoka, M. Ban, and M. Sasaki, Phys. Rev. A 68,
012307 (2003).\\
A. Chefles, Phys. Lett. A 239, 339 (1998)\\
D. Dieks, Phys. Lett. A 126, 303 (1998)\\
A. Peres, Phys. Lett. A 128, 19 (1988)\\
A. Chefles and S. M. Barnett, Phys. Lett. A 250, 223
(1998)\\
Y. Sun, M. Hillery, and J. A. Bergou, Phys. Rev. A 64,
022311 (2001)\\
J. A. Bergou, M. Hillery, and Y. Sun, J. Mod. Opt. 47, 487
(2000)\\
Y. Sun, J. A. Bergou, and M. Hillery, Phys. Rev. A 66,
032315 (2002)\\
J. A. Bergou, U. Herzog, and M. Hillery, Phys. Rev. Lett.
90, 257901 (2003)\\
M. Mohseni, A.M. Steinberg, and J. Bergou, Phys. Rev. Lett. 93, 200403 (2004)\\
M.A.P. Touzel, R.B.A.Adamson, and A.M. Steinberg, Phys. Rev. A 76, 062314 (2007)\\

\vspace{0.2in}\\

{\bf Complementarity and quantum erasers:}\\

See, e.g., Z.Y. Ou, L.J. Wang, X.Y. Zou, and L. Mandel, Phys Rev A 41, 566 (1990)\\
M. Hillery and M.O. Scully, in Quantum Optics, Experimental Gravitation, and Measurement Theory, edited by P. Meystre {\it et al.} (Plenum: New York, 1983), pp. 65-85 \\
M.O. Scully, B.-G. Englert, and H. Walther, Nature 351, 111 (1991)\\
D.M. Greenberger and A. Yasin, Phys. Lett. A 128, 391 (1988)\\
G. Jaeger, A. Shimony, and L. Vaidman, Phys. Rev. A 51, 54 (1995)\\
B.-G. Englert, Phys. Rev. Lett. 77, 2154 (1996)\\
P.G. Kwiat, A.M. Steinberg, and R.Y. Chiao, Phys. Rev. A 45, 7729 (1992)\\
Y.-H. Kim, R. Yu, S.P. Kulik, Y.H. Shih, and M.O. Scully, Phys. Rev. Lett. 84, 1 (2000)\\

For a review, see A.M. Steinberg, P.G. Kwiat, and R.Y. Chiao, AMO Physics Handbook (AIP Press, edited by GWF Drake, 1996); available online at http://www.physics.utoronto.ca/$\sim$steinber/Quantum\_Optical.pdf\\

\vspace{0.2in}\\

{\bf Interaction-free measurements:}\\
A.C. Elitzur and L. Vaidman, Found. Phys. 23, 987 (1993)\\
P.G. Kwiat, H. Weinfurter, T. Herzog, A. Zeilinger and
M.A. Kasevich, Phys. Rev. Lett. 74, 4763(1995)\\
L. Hardy, Phys. Rev. Lett. 68, 2981 (1992)\\
Y. Aharonov et al., Phys. Lett. A 301, 130 (2002)\\
J.S. Lundeen and A.M. Steinberg, Phys. Rev. Lett 102, 020404 (2009)\\
K. Yokota, T. Yamamoto, M. Koashi, and N. Imoto, New. J. Phys. 11, 033011 (2009)\\

\vspace{0.2in}\\

{\bf Cloning:}\\
W.K. Wootters and W.H. Zurek, Nature 299, 802 (1982)\\
A. Peres, ``How the no-cloning theorem got its name,'' quant-ph/0205076 (2002)\\
N. Herbert. Found. Phys. 12, 117 (1982) \\
P.W. Milonni and M.L. Hardies. Phys. Lett. 92A, 321 (1982)\\
A. Garuccio, in The Present Status of the Quantum Theory of 
	Light, S. Jeffers et al. ed's, Kluwer (Dordrecht: 1997)\\
K. Furuya, P.W. Milonni, A.M. Steinberg, and M. Wolinsky, Phys. Lett. A 251, 294 (1999)\\
E. Nagali, T. de Angelis, F. Sciarrino, and F. de Martini, Phys. Rev. A 76, 042126 (2007) \\
J. Fiur{\'a}sek and Cerf, Phys. Rev. A 77, 052308 (2008)\\
J.-S. Xu, C.-F. Li, L. Chen, X.-B. Zou, and G.-C. Guo, Phys. Rev. A 78, 032322 (2008)\\

\vspace{0.2in}\\

{\bf Dense coding and teleportation:}\\

C.H. Bennett and S.J.Wiesner, Phys. Rev. Lett. 69, 2881 (1992)\\
K. Mattle, H. Weinfurter, P.G. Kwiat, and A. Zeilinger, Phys. Rev. Lett.  76, 4656 (1996)\\
C.H. Bennett, G. Brassard, C. Cr{\'e}peau, R. Jozsa, A. Peres, and W.K. Wootters, Phys. Rev. Lett.  70, 1895 (1993)\\
D. Bouwmeester, J.-W. Pan, K. Mattle, M. Eibl, H. Weinfurter, and A. Zeilinger, Nature 390, 575 (1997)\\
A. Furusawa, J.L. S{\o}rensen, S.L. Braunstein, C.A. Fuchs, H.J. Kimble, and E.S. Polzik,  Science 282, 706 (1998)\\

\vspace{0.2in}\\

{\bf Error-correcting codes:}\\
A. Steane, Proc. Roy. Soc. Lond. A 452, 2551 (1996)\\
P.W. Shor, Phys. Rev. A 52, R2493 (1995)\\
E. Knill, R. Laflamme, A. Ashikhmin, H. Barnum, L. Viola, and W.H. Zurek, quant-ph/0207170 (2002)\\

\vspace{0.2in}\\

{\bf Single-photon  tomography:}\\
A.G. White, D.F.V. James, W.J. Munro, and P.G. Kwiat, Phys. Rev. A 65, 012301 (2001)\\
D.F.V. James, P.G. Kwiat, W.J. Munro, and A.G. White, Phys. Rev. A 64, 052312 (2001)\\
{\bf Ancilla-assisted photon-polarisation tomography:}\\
J. Altepeter et al., Phys. Rev. Lett.  90, 193601 (2003)\\
{\bf Phase-space tomography on single-photon fields:}\\
A.I. Lvovsky et al., Phys. Rev. Lett.  87, 050402 (2001)\\
{\bf Two-photon process tomography:}\\
M.W. Mitchell et al., Phys. Rev. Lett. 91, 120402 (2003)\\
{\bf Applications of process tomography:}
Y.S. Weinstein et al., Phys. Rev. Lett.  86, 1889 (2001) \\
N. Boulant, M.A. Pravia, E.M. Fortunato, T.F. Havel, and D.G. Cory, QIP 1, 135 (2002) [see also quant-ph/0211046]\\
A.G. White et al., J. Opt. Soc. Am. B 24, 172 (2007)\\

\vspace{0.2in}\\

{\bf Linear-optics quantum computation:}\\
E. Knill, R. Laflamme, and G.J. Milburn, Nature 409, 46 (2001)\\
D. Gottesmann and I.L. Chuang, Nature 402, 390 (1999)\\
T.C. Ralph, N.K. Langford, T.B. Bell, and A.G. White, Phys. Rev. A 65, 062324 (2002)\\
T.B. Pittman, B.C. Jacobs, and J.D. Franson, Phys. Rev. Lett. 88, 257902 (2002)\\
J.L. O'Brien, G.J. Pryde, A.G. White, T.C. Ralph, and D. Branning, Nature 426, 264
(2003)\\
N.K. Langford et al., Phys. Rev. Lett. 95, 210504 (2005)\\

\vspace{0.2in}\\

{\bf Measurement-based quantum computation and related approaches}\\
M.A. Nielsen, "Universal quantum computation using only\ldots", quant-ph/0108020; see also Phys. Lett. A 308, 96 (2003)\\
R. Raussendorf and H.J. Briegel, Phys. Rev. Lett. 86, 5188 (2001)\\
R. Raussendorf and H.J. Briegel, Phys. Rev. A 68, 022312 (2003)\\
P. Aliferis and D.W. Leung, Phys. Rev. A 70, 062314 (2004)\\
M.A. Nielsen, Rep. Math. Phys. 57, 147 (2006)\\
P. Walther et al, Nature 434, 169 (2005)\\
H.J. Briegel et al, Nature Physics 19 (2009) \\
M. Anderlini et al, Nature 448, 452 (2007) \\
K. Nemoto and W.J. Munro, Phys. Rev. Lett 93, 250502 (2004) \\
J.D. Franson, B.C. Jacobs, and T.B. Pittman, Phys. Rev. A 70, 062302 (2004)\\

\vspace{0.2in}\\

{\bf N00N states and generation:}:
H. Lee et al., Phys. Rev. A 65, 030101 (2002)\\
J. Fiur{\'a}sek, Phys. Rev. A 65, 053818 (2002)\\
M.W. Mitchell et al., Nature 429, 161 (2004)\\
K.J. Resch et al., Phys. Rev. Lett. 98, 223601 (2007)\\

\vspace{0.2in}\\

{\bf Weak measurements:}
Y. Aharonov and L. Vaidman, Phys. Rev. A 41, 11 (1991)\\
Y. Aharonov, D.Z. Albert, and L. Vaidman, Phys. Rev. Lett.  60, 1351 (1988)\\
N.W.M. Ritchie, J.G. Story, and R.G. Hulet, Phys. Rev. Lett.  66, 1107 (1991)\\
A.M. Steinberg, Phys. Rev. A 52, 32 (1995)\\
H.M. Wiseman, Phys. Rev. A 65, 032111 (2002)\\
N. Brunner, A. Ac\`in, D. Collins, N. Gisin, and V. Scarani, Phys. Rev. Lett. 91, 180402 (2003)\\
K.J. Resch and A.M. Steinberg, Phys. Rev. Lett.  92, 130402 (2004)\\
R. Mir et al., New J. Phys. 9, 287 (2007)\\
O. Hosten and P.G. Kwiat, Science 319, 787 (2008)\\
J.S. Lundeen and A.M. Steinberg, Phys. Rev. Lett.  102, 020404 (2009)\\
K. Yokota, T. Yamamoto, M. Koashi, and N. Imoto, New. J. Phys. 11, 033011 (2009)\\
P. Ben Dixon, D.J. Starling, A.N. Jordan, and J.C. Howell, Phys. Rev. Lett.  102, 173601 (2009)\\
S. Kocsis et al., Science 332, 1170 (2011)\\
A. Feizpour, X. Xing, and A.M. Steinberg, Phys. Rev. Lett.  107, 133603 (2011)\\

\endthebibliography

%

\end{document}